\documentclass{aa}   % two columns

\usepackage{graphicx}
\usepackage{txfonts}
\usepackage{hyperref} 	
\usepackage{color}
\usepackage{subfig}

\newcommand{\hii}{\hbox{H\,{\sc ii}}}
\newcommand{\hi}{\hbox{H\,{\sc i}}}

\newcommand{\oiid}{[\ion{O}{ii}]$\lambda\lambda3726,3729$}

\newcommand{\oii}{[\ion{O}{ii}]}
\newcommand{\oiii}{[\ion{O}{iii}]}
\newcommand{\oiiiopt}{[\ion{O}{iii}]$\lambda4959$}
\newcommand{\oiiibpt}{[\ion{O}{iii}]$\lambda5007$}

\newcommand{\siid}{[\ion{S}{ii}]$\lambda\lambda6717,6731$}

\newcommand{\civd}{\ion{C}{iv}$\lambda\lambda1548,1551$}

\newcommand{\ciiid}{[\ion{C}{iii}]$\lambda1907$+\ion{C}{iii}]$\lambda1909$}
\newcommand{\ciiit}{\ion{C}{iii}]$\lambda1908$}
\newcommand{\ciii}{\ion{C}{iii}]}

\newcommand{\heii}{\ion{He}{ii}$\lambda1640$}

\newcommand{\neiiimuse}{[\ion{Ne}{iii}]$\lambda3869$}
\newcommand{\neiii}{[\ion{Ne}{iii}]}

\newcommand{\mgiit}{\ion{Mg}{ii}$\,\lambda\lambda2796,2803$}
\newcommand{\mgii}{\ion{Mg}{ii}}
\newcommand{\mgi}{\ion{Mg}{i}}
\newcommand{\mgiib}{\ion{Mg}{ii}$\lambda2796$}
\newcommand{\mgiir}{\ion{Mg}{ii}$\lambda2803$}

\newcommand{\cloudy}{\textsc{cloudy}}

\newcommand{\beagle}{\textsc{beagle}}
\newcommand{\galfit}{\texttt{GALFIT}}
\newcommand{\platefit}{\texttt{PLATEFIT}}

\newcommand{\mgiisample}{\mgii\ parent sample}
\newcommand{\udften}{\textsf{udf10}}
\newcommand{\mosaic}{\textsf{mosaic}}

\newcommand{\msun}{M$_{\odot}$}
\newcommand{\zsun}{$Z_{\odot}$}

\newcommand{\logU}{log $\langle$U$\rangle$}

\begin{document}

\title{The MUSE Hubble Ultra Deep Field Survey}
\subtitle{XII. \ion{Mg}{II} emission and absorption in star-forming galaxies \thanks{Based on observations made with ESO telescopes at the La Silla
Paranal Observatory under programs 094.A-0289(B), 095.A-0010(A),
096.A-0045(A) and 096.A-0045(B).}}
    \titlerunning{\ion{Mg}{II} emission and absorption in star-forming galaxies}
    
\author{Anna Feltre\inst{1} \thanks{\email{anna.feltre@univ-lyon1.fr}}
  	\and 
  	Roland Bacon\inst{1}
 	\and 
	Laurence Tresse\inst{1}
	\and
	Hayley Finley\inst{2,3}
	\and
	David Carton\inst{1}
	\and
	J\'er\'emy Blaizot\inst{1}
	\and
	Nicolas Bouch\'e\inst{2}
	\and 
	Thibault Garel\inst{1}	
	\and
	Hanae Inami\inst{1}
	\and
	Leindert A. Boogaard\inst{4}
	\and
	Jarle Brinchmann\inst{4,5}
	\and
	St\'ephane Charlot\inst{6}
	\and
	Jacopo Chevallard\inst{6}
	\and
	Thierry Contini\inst{2}
	\and
	Leo Michel-Dansac\inst{1}
	\and
	Guillaume Mahler\inst{1,7}
	\and
	Raffaella A. Marino\inst{8}
	\and
	Michael V. Maseda\inst{4}
	\and
	Johan Richard\inst{1}
	\and
	Kasper B. Schmidt\inst{9}
	\and
	Anne Verhamme\inst{1,10}}

\institute{Univ Lyon, Univ Lyon1, Ens de Lyon, CNRS, Centre de Recherche Astrophysique de Lyon UMR5574, F-69230, Saint-Genis-
Laval, France   \email{anna.feltre@univ-lyon1.fr}
	\and
	Institut de Recherche en Astrophysique et Plan\'etologie (IRAP), Universit\'e de Toulouse, CNRS, UPS, F-31400 Toulouse, France
	\and
	Stockholm University, Department of Astronomy and Oskar Klein Centre for Cosmoparticle Physics, AlbaNova University Centre, SE-10691, Stockholm, Sweden
	\and
	Leiden Observatory, Leiden University, P.O. Box 9513, 2300 RA Leiden, The Netherlands
	\and 
	Instituto de Astrof{\'\i}sica e Ci{\^e}ncias do Espa\c{c}o, Universidade do Porto, CAUP, Rua das Estrelas, PT4150-762 Porto, Portugal
	\and
	Sorbonne Universit\'es, UPMC-CNRS, UMR7095, Institut d'Astrophysique de Paris, F-75014, Paris, France
	\and
	Department of Astronomy, University of Michigan, 1085 South University Ave, Ann Arbor, MI 48109, USA
	\and
	Department of Physics, ETH Z\"urich, Wolfgang-Pauli-Str. 27, 8093 Z\"urich, Switzerland
	\and
	Leibniz-Institut f\"ur Astrophysik Potsdam (AIP), An der Sternwarte 16, 14482 Potsdam, Germany
	\and
	Observatoire de Gen\`eve, Universit\'e de Gen\`eve, 51 Ch. des Maillettes, 1290 Versoix, Switzerland
	}

\date{Received X Month XXXX / Accepted X Month XXXX}

\abstract{

The physical origin of the near-ultraviolet \mgii\ emission remains an under-explored domain, 
contrary to more typical emission lines detected in the spectra of star-forming galaxies.
We explore the nebular and physical properties for a sample of 381 galaxies between 0.70 < z < 2.34 
drawn from the MUSE {\it Hubble} Ultra Deep Survey.
The spectra of these galaxies show a wide variety of profiles of the \mgiit\ resonant doublet, from absorption to emission. 

We present a study on the main drivers for the detection of \mgii\ emission in galaxy spectra.
By exploiting photoionization models we verified that the emission-line ratios observed in galaxies with \mgii\ in emission
are consistent with nebular emission from H{\sc ii} regions. 
From a simultaneous analysis of MUSE spectra and ancillary HST information
via spectral energy distribution (SED) fitting, we find that galaxies with \mgii\ in emission have lower stellar masses, smaller sizes, bluer spectral slopes and lower optical depth than those with absorption.
This leads us to suggest that \mgii\ emission is a potential tracer of physical conditions not merely related to those of the ionized gas.
We show that these differences in \mgii\ emission/absorption can be explained in terms of a higher dust and neutral gas content in the interstellar medium (ISM) of galaxies showing \mgii\ in absorption,
confirming the extreme sensitivity of \mgii\ to the presence of the neutral ISM. 

We conclude with an analogy between the \mgii\ doublet and the Ly$\alpha$ line, due 
to their resonant nature. Further investigations with current and future facilities, including JWST, are promising as 
the detection of \mgii\ emission and its potential connection with Ly$\alpha$ could provide new insights on the ISM content in the early Universe.
}

  
\maketitle

\section{Introduction}\label{sec:intro}

Interpreting the physical nature of the spectral features observed in galaxy spectra is a non trivial path to understand the physical processes
at work within galaxies and, through that, the galaxy population evolution through cosmic time. 
While optical lines have been extensively studied, ultraviolet (UV) lines are recently under scrutiny (see \citealt{Stark2016} for an exhaustive review). 
The most explored ones are the Lyman-$\alpha \, \lambda1215.67$ (hereafter Ly$\alpha$) line, 
along with the \ciiid\ (hereafter \ciii) emission doublet, detected in the redshifted spectra of distant galaxies \citep[e.g.][]{Stark2015a,Maseda2017,Nakajima2018a}.
In addition, the complex profiles of combined stellar and nebular \civd\ and \heii\ emissions are observed both 
in the spectra of local metal-poor galaxies \citep[e.g.][]{Berg2016,Senchyna2017} and higher redshifts gravitationally-lensed galaxies 
\citep[e.g.][and references therein]{Stark2015b,Vanzella2016,Mainali2017,Berg2018}.

The near-UV \mgiit\ resonant doublet (hereafter \mgii) has been detected (either in emission and/or absorption)
both in planetary nebulae and galaxy spectra. 
Intriguing, early works on planetary nebulae found the \mgii\ doublet to be absent despite
the presence of fainter Magnesium emission lines, such as \ion{Mg}{i}$\lambda4572$, 
\ion{Mg}{i}$\lambda4562$, \ion{Mg}{i}$\lambda4481$ and \ion{Mg}{ii}$\lambda4391$ in the same spectra. 
This led to the conclusion that \mgii\ was likely an extremely sensitive tracer 
of some specific physical conditions of the gaseous nebulae themselves \citep{Gurzadyan1997}.

The first detection of \mgii\ emission in a starbursting galaxy comes from the IUE ({\itshape International Ultraviolet Explorer}) 
spectrum of Tol1924-416 \citep{Kinney1993}. 
In the last decade, the redshifted \mgii\ emission has been detected in several studies of galactic winds 
\citep{Weiner2009,Rubin2010,Rubin2011,Giavalisco2011,Martin2012,Erb2012,Kornei2013,Finley2017}
and in the spectra of gravitationally-lensed galaxies \citep{Rigby2014,Karman2016,Bordoloi2016}.
This feature is often accompanied by blueshifted absorption, yielding a profile similar to a P-Cygni feature. 
Several explanations have been proposed for the origin of the \mgii\ emission, including the presence of an 
active galactic nucleus, AGN \citep{Weiner2009}, and resonant scattering in expanding winds \citep{Rubin2010,Erb2012}. 

\cite{Erb2012} studied large-scale outflows in a sample of 96 star-forming galaxies at $1\lesssim z\lesssim2$ with the \mgii\ 
doublet ranging from emission to absorption. They found
\mgii\ emission to be more common both at lower stellar masses and in galaxies with bluer UV slopes. 
\cite{Kornei2013} detected \mgii\ emission in $\sim$15\% of a sample of 212 star-forming galaxies at $z\sim1$, selected from the DEEP2 survey \citep{Newman2013}.
They found that these sources had higher specific star formation, lower dust attenuation and lower stellar masses compared to the whole sample. 
\cite{Guseva2013} detected the \mgii\ doublet emission in 45 low-metallicity star-forming galaxies within $0.36<z<0.7$ from a sample of
62 from the Sloan Digital Sky Survey, SDSS \citep{York2000} and determined a Magnesium over Oxygen abundance ratio a factor $\sim2$ lower than the solar one.
These studies showed that the detection of the \mgii\ emission feature 
is not limited to rare peculiar sources, as thought after its first detections, but concerns a significant
fraction of objects within different galaxy samples \citep[e.g.][]{Erb2012, Guseva2013}.
Indeed, \mgii\ emission, either in pure emission or P-cygni profiles, has also been detected with the Multi-Unit Spectroscopic Explorer \citep[MUSE][]{Bacon2015} 
in 50 star-forming galaxies from a sample of 271 \oiid\ emitters \citep{Finley2017}.
The different profiles observed in these galaxies contain valuable clues on the physical origin of 
\mgii\ emission. Once AGN are excluded, \mgii\ emission might originate from nebular emission in H{\sc ii} regions, 
with subsequent resonant scattering in neutral (or low ionization) gas, and/or resonant scattering of continuum photons in outflowing gas. 
Which of these is the dominant physical process for a given \mgii\ profile is still unclear.

\cite{Rigby2014} examined the \mgii\ P-Cygni profiles observed in the spectra of five gravitationally-lensed bright star-forming galaxies ($1.66 < z < 1.91$),
along with other spectral features, including Ly$\alpha$.
Given that \mgii\ and Ly$\alpha$ are both resonantly scattered lines, their physics is analogous. Provided that the lines are produced by the same mechanism 
and observed through the same gas, one would expect their observed properties to be correlated.
However, \cite{Rigby2014} found a lack of correlation between \mgii\ (in P-cygni profile) and Ly$\alpha$, suggesting reprocessed stellar continuum, as responsible for the bulk of \mgii\ emission.
Very recently, \cite{Henry2018} found a close relation between the Ly$\alpha$ and \mgii\ profiles in a sample 10 Green Pea galaxies at $z\sim0.2-0.3$.
They also found \mgii\ emission to be associated to low, if not null, dust absorption.

No less important is the study of the asymmetric, blue-shifted \mgii\ absorption profile, commonly used to identify outflowing gas within galaxies up to z$\sim$2 \citep[e.g.][in addition to those mentioned above]{Veilleux2005, Tremonti2007,Steidel2010,Harikane2014,Zhu2015,Finley2017a}.
Indeed, both models and observations support the importance of galactic winds in regulating the metal enrichment of the intergalactic medium and the chemical evolution of galaxies 
\citep[e.g.][]{Aguirre2001,Tremonti2004,Finlator2008,Mannucci2009,Lilly2013}.

MUSE enabled the detection of a large number of \mgii\ emitters, along with absorbers and P-Cygni, for a relatively 
large redshift range ($0.70 \leq z \leq 2.34$).
These spectra provide valuable clues on the excitation properties of these sources, 
thanks to the additional emission lines detected in their spectra, such as \oiid, \neiiimuse\ (hereafter \oii and \neiii) and \ciiit.
The additional availability of broad-band photometry from HST 
allows a multi-band coverage from UV to near-infrared continuum.  

Here, we assemble 381 galaxies 
from the MUSE {\it Hubble} Ultra Deep Survey \citep{Bacon2017} 
in the redshift range $0.70 \leq z \leq 2.34$, covering the peak of the star formation rate density \citep[SFRD,][]{Madau2014}.
Our aim is to further explore the variety of profiles shown by the \mgii\ doublet (emission, P-Cygni, absorption)
and to understand what is the main driver for this variety.
The main goal is to investigate whether the galaxy properties differ in terms of
 metallicity, ionization parameter, stellar mass, star formation rate (SFR) and dust attenuation.
 We focus on the differences between galaxies showing \mgii\ in emission and those with \mgii\ in absorption, leaving a
detailed study of the sources showing \mgii\ P-Cygni profile to future works.
We infer the physical properties of our galaxies by exploiting the synergy between MUSE and HST, 
through the combined use of newly developed photoionization models \citep{Gutkin2016} 
and the Bayesian statistics fitting tool BEAGLE  \citep{Chevallard2016}.

The paper is structured as follows: Sect. \ref{sec:sample} describes the sample selection and classification, 
along with the observed properties from HST photometry and MUSE spectra. 
Sect. \ref{sec:nebular} investigates how the \mgii\ emission features
compare with predictions from photoionization models. 
The description of the spectral fitting technique and the main results from the analysis are 
described in Sect. \ref{sec:fitting} and are followed by discussions and conclusions in Sect. \ref{sec:discussion} and \ref{sec:conclusions}, respectively. 
Throughout the paper, we use the AB flux normalisation, we follow a convention
where negative/positive equivalent widths (EW) correspond to emission/absorption
and we adopt the cosmological parameters
from \cite{Planck2016}, ($\Omega_{\rm M}$, $\Omega_{\lambda}$, $H_{\rm 0}$) = (0.308, 0.692, 67.81).

%--------------------------------------------------------------------------------------------------------------------------
\section{The \mgii\ sample}\label{sec:sample_selection}

\subsection{MUSE Observations and Spectral Measurements}\label{sec:catalogue}

We assembled a sample of galaxies drawn from the MUSE {\it Hubble} Ultra Deep Field Survey \citep[][]{Bacon2017}.
This two-layered spectroscopic survey covers 90\% of the total {\itshape Hubble} Ultra Deep Field (HUDF)
and comprises a ($3' \times 3'$) mosaic of nine MUSE fields (hereafter \mosaic) with an exposure time of 10-hours, 
in addition to a deeper exposure of 31-hours in a single field (hereafter \udften) of 1.15 arcmin$^{2}$.  
The 50\% spectroscopic completeness in the HST/F775W wide-band is reached at 26.5 mag for \udften\ and 25.5 mag for the \mosaic\ \citep{Inami2017}.
Redshifts and line flux measurements from the first Data Release (DR1) of the HUDF survey are described in \cite{Inami2017}. 
The DR1 catalogue includes 1338 sources with a MUSE-based redshift confidence level above or equal to 2 \citep[Sec 3.2 of][]{Inami2017}, 
of which 253 lie in the \udften\ field.

The line intensities in the DR1 catalogue are computed on unweighted summed spectral extractions of the data cube \citep[sections 3.1.3 and 3.3 of][]{Inami2017}
using the \platefit\ software \citep{Brinchmann2004,Tremonti2004}. \platefit\ fits the continuum of the observed spectrum, where strong emission lines have been masked out, 
 with a set of theoretical templates from \cite{Bruzual&Charlot2003} computed using the MILES \citep{Sanchez-Blazquez2006} stellar spectra. 
 After the continuum subtraction, the procedure simultaneously fits a single Gaussian profile to each expected emission line.
The line fluxes and EW used in Sec. \ref{sec:nebular}, \ref{sec:results} and \ref{sec:fesc} have been computed with the same procedure as for the MUSE HUDF DR1 catalogue
but using weighted optimal spectral extractions, namely white-light weighted and PSF weighted, accordingly to the weighted spectra used to measure the systemic redshfit
(identified with \texttt{REF$\_$SPEC} in the HUDF DR1 catalogue). The advantages of using weighted extractions are an higher signal-to-noise ratio (S/N) and
a reduced contamination from neighbouring sources.
We have, however, checked that using unweighted summed spectral extractions does not change the conclusions of this analysis.

We note that the line fluxes and EW of \mgii\ have been recomputed with a different \platefit\ setup allowing for a 
potential velocity difference between the \mgii\ resonant transition and the systemic redshift.
This is necessary because, as described later in Sect. \ref{sec:em_lines}, the DR1 catalogue assumes that all emission lines have the same
intrinsic velocity shift relative to the systemic redshift \citep[see section 3.3 of][for details]{Inami2017}.
However, since \mgii\ is a resonant line, it might have a different velocity shift and width, leading us to underestimate its intensity.
  
\subsection{Sample selection}\label{sec:sample}

From the combined catalogue of the \udften\ and \mosaic\ fields (excluding duplicates) 
we selected sources with a spectroscopic redshift $0.70 \leq z \leq 2.34$, 
to ensure MUSE spectral coverage of the \mgiit\ doublet wavelengths.
We additionally required the redshift to be measured with a confidence level \texttt{CONFID} $>1$ \citep[see section 3.2. of][]{Inami2017}, 
finding 403 sources satisfying the above requirements.
As the process for the systemic redshift determination of the DR1 does not include the resonant \mgii\ doublet, 
this cut secures at least one spectral feature in the MUSE spectra, regardless of the detection of an \mgii\ spectral feature. 

The above selection criteria include the \mosaic\ source ID 872, 
which is an AGN showing a prominent and broad \mgii\ emission feature \citep[see also][]{Inami2017}
and has been excluded from the following analysis. 
In addition, as explained in Sect. \ref{sec:others}, we discarded 10 sources classified as AGN on the basis of their X-ray spectra from the 
7 Ms Source Catalogs of the Chandra Deep Field-South Survey \citep[][see their section 4.5. for source classification]{Luo2017}.
Moreover, we removed 11 sources that do not have HST broad-band photometry available (Sec. \ref{sec:hst}), because of the ambiguity in associating the HST counterpart.
The final parent sample has 381 galaxies, 63 in the \udften\ and 318 in the \mosaic-only fields. 

\subsection{Sample classification}\label{sec:classification}

\begin{figure*}
  \centering
  \includegraphics[width=15cm]{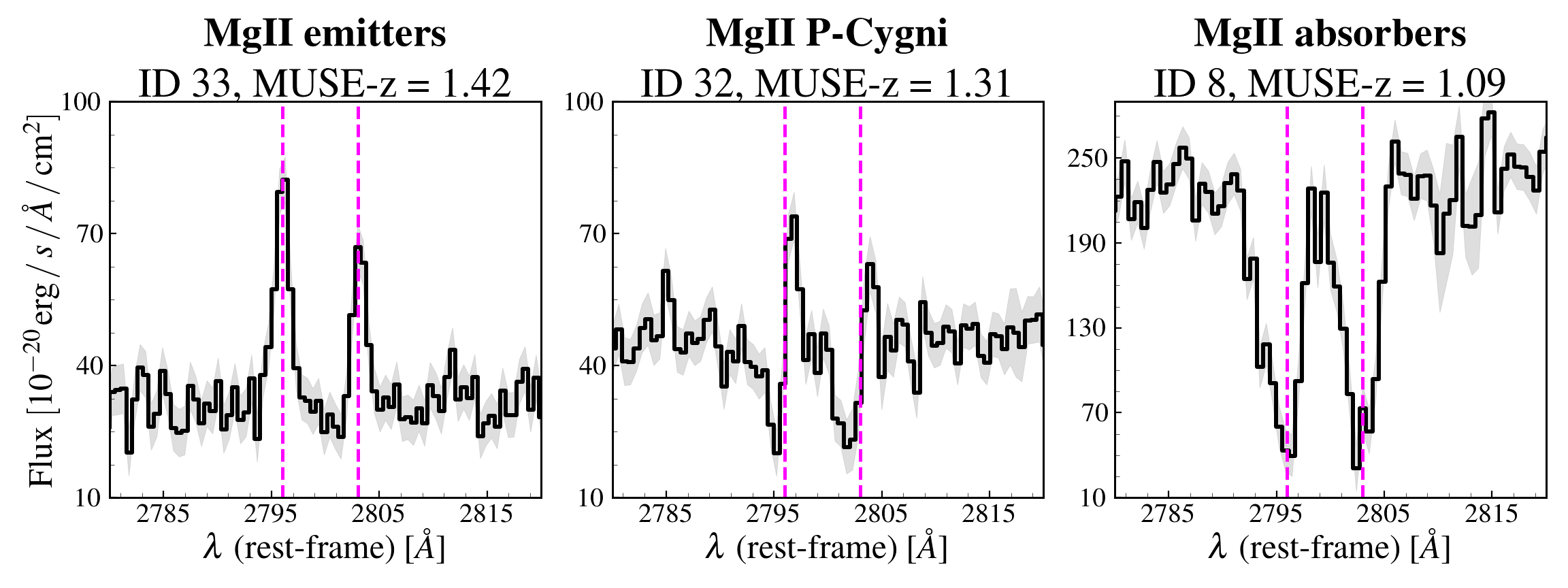}
   \caption{From left to right: zoom-in of MUSE spectra, at rest-frame wavelengths,  
   of sources showing \mgii\ in emission, P-Cygni profile and absorption. 
   Magenta dashed lines indicate the rest-frame wavelengths of the \mgiib\ and \mgiir\ doublet components.}
            \label{Fig1}%
   \end{figure*}

The \mgii\ doublet in our spectra shows a wide variety of profiles, ranging from clear emission, 
to blueshifted absorption with redshifted emission (P-Cygni-like profiles), to strong deep absorption, as illustrated in Fig. \ref{Fig1}.
We classified the sources in four spectral types, namely emitter, P-Cygni, absorber and non-detection,
on the basis of both the intensity of the \mgii\ profile and quality of the spectrum, as follows:

\begin{itemize}
\item {\bf Mg\,{\sc ii} emitters} 
\begin{itemize}
\item both components of the doublet showing emission EW \mgii\ $< -1$;
\item good S/N ($> 3$) from \platefit\ in either both components of the \mgii\ doublet or the strongest (see Sec. \ref{sec:emitters}) \mgiib\ component;
\end{itemize}
\item {\bf Mg\,{\sc ii} P-Cygni} 
\begin{itemize}
\item profiles have been visually inspected;
\end{itemize}
\item {\bf Mg\,{\sc ii} absorbers}
\begin{itemize}
 \item EW \mgii\ $> +1$;
 \item MUSE spectrum with a ${\rm S/N} >3$, averaged in a window of $30\,\AA$ centred at $2800\,\AA$;
\end{itemize}
\item {\bf Mg\,{\sc ii} non-detection}
\begin{itemize} 
\item i.e. all the remaining sources in the \mgii\ parent sample.
\end{itemize}
\end{itemize}

\begin{figure}
   \hspace*{-0.7cm}  
  \includegraphics[width=10.5cm]{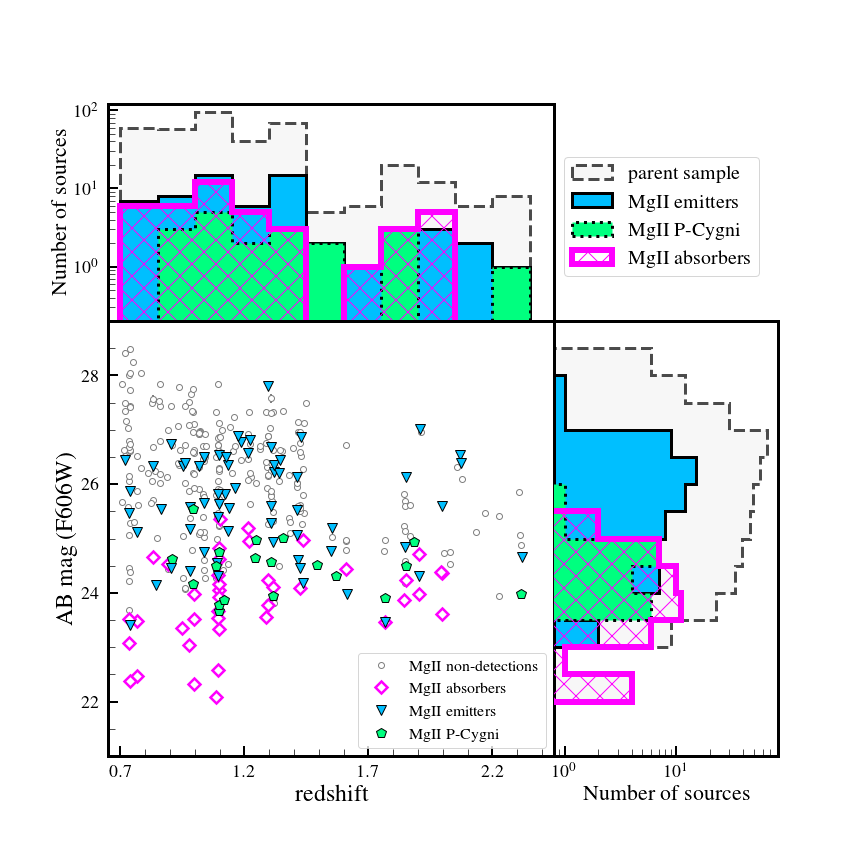}
   \caption{Redshift (top panel) and F606W HST passband filter flux (right panel) distributions for the whole \mgiisample\ (dark gray dashed-line histrogram). 
   As labeled in the legend, distributions of \mgii\ emitters, P-Cygni and absorbers
   are shown in cyan, green and magenta histograms, respectively.}
            \label{Fig2}%
   \end{figure}

As summarised in Table \ref{table1}, 
the \mgiisample\ consists of 63/19/41/258 \mgii\ emitters/P-Cygni/absorbers/non-detections, respectively. 
If the ${\rm S/N}$ criteria to detect \mgii\ emitters were reduced to 2 we would have obtained 33 additional sources.
In 12 of these 33 galaxies, the \mgii\ line fall in spectral region redward 7500 \AA\ where the flux uncertainties are larger because of the strong skyline contamination. 
Instead, relaxing the EW threshold to $EW<-0.5$ would have given us 3 additional sources with ${\rm S/N}>3$.
We adopt a $S/N>3$ and $EW<-1.0$ to avoid including possible contaminants in the sample.
Note that, we further validated and refined this classification with a thorough visual inspection. 
The \mgiisample\ has 261 galaxies in common with the sample of \oii\ emitters from \cite{Finley2017}, selected
with the aim of studying non-resonant \ion{Fe}{ii}*($\lambda2365$, $\lambda2396$, $\lambda2612$, $\lambda2626$)
transitions as potential tracer of galactic outflows. \cite{Finley2017} have visually inspected the spectra of their galaxies and
flagged the \mgii\ profile of their sources in pure emission, P-Cygni and pure absorption.
We verified that the classifications of the galaxies in common between the two samples are in good overall agreement. 
The classification of a galaxy as \mgii\ P-Cygni depends on our ability to detect the blueshifted absorption and, hence, both on spectral noise and resolution.
Our \mgii\ P-Cygni have a spectrum with ${\rm S/N}>3$ in a window of $30\,\AA$ centred at $2800\,\AA$ and their faintest HST flux in the F606W passband filter is 25.5 mag 
(see also Fig. \ref{Fig2}). More quantitative measurements on the amount of absorption and emission in \mgii\ P-Cygni sources, and corresponding EW measurements, will be provided in Finley et al., in prep.

17\% of the galaxies in the \mgiisample\ are classified as \mgii\ emitters, similar to the $\sim$15\% of \cite{Kornei2013}.
This fraction differs from $\sim$1/3 of \mgii\ emitters detected in the sample of \cite{Erb2012} at z$\sim$2 and
from the $\sim2/3$ of the low redshift SDSS galaxy sample of \cite{Guseva2013}.
However, the different fractions of \mgii\ emitters among the samples may be related to the different selection and classification criteria. 
The \cite{Erb2012} sample was photometrically 
preselected in the rest-UV, while the galaxies of \cite{Guseva2013} were selected to
have low-metallicity H{\sc ii} regions with strong emission lines.
In our work, we do not apply any pre-selection but simply explore the \mgii\ profiles in all the galaxies
for which we had MUSE spectral coverage of the \mgiit\ wavelengths 
(upon selecting sources with good redshift measurement and minimizing the AGN contribution, see Sect. \ref{sec:sample}). 
Moreover, a blueshifted absorption, tracer of stellar winds, accompanying the \mgii\ redshifted emission,
was detected in many spectra of the \mgii\ emitters in both the \cite{Erb2012} and \cite{Guseva2013} samples. 
In this work, sources with this profile are classified as \mgii\ P-Cygni and treated separately.

\begin{table}
\caption{Classification of the \mgiisample}              
\label{table1}      
\centering                                      
\begin{tabular}{ c | c | c | c | c }          
\hline\hline                        
Field & Emitters & P-Cygni & Absorbers & Non-detections \\   
\hline\hline                                    
    \sf{udf-10} & 18 & 3 & 11 & 31\\      
        \hline  
    \sf{mosaic} & 45 & 16 & 30 & 227\\
    \hline  
     combined & 63 & 19 & 41 & 258\\
\hline            
\end{tabular}
\end{table}

%-----------------------------
% Figure 1

The redshift distribution of the \mgiisample\ ($0.7\leq z \leq 2.34$) is shown in Fig. \ref{Fig2} (top panel). 
We found no particular redshift preference for the occurrence of \mgii\ emitters 
and absorbers compared to the whole sample. 
The p-value of 0.3 from a two-sample Kolmogorov-Smirnov (KS) test supports
a similar redshift distribution for the two spectral types of galaxies (\mgii\ emitters and absorbers).
The same is true for the sources with \mgii\ P-Cygni profile.

%-----------------------------
% Figure 2

\subsection{Observational properties from HST photometry and imaging}\label{sec:hst}

HST broad-band photometry from UVUDF \citep[11 HST/WFC3 and ACS photometric bands,][]{Rafelski2015} is available for the whole parent sample
and probes the \mgiit\ wavelengths with the F606W, F775W and F850LP passband filters 
for sources within the redshift intervals of $0.65 \lesssim z \lesssim 1.56$, $1.43 \lesssim z \lesssim 2.0$ and $1.86 \lesssim z \lesssim  2.34$ , respectively. 
The observed luminosities of the \mgii\ sample range from 22.08 (21.59, 20.73) to 29.87 (29.81, 30.16) mag in the F606W (F775W, F850LP) HST filters, respectively.
Fig. \ref{Fig2} (right panel) shows the F606W passband filter distribution for the whole \mgiisample.
We note that 20\% of the \mgii\ non-detections are among the faintest (F606W flux $\gtrsim$ 27 mag) galaxies in the sample, while all but one 
of the \mgii\ emitters have F606W flux brighter than 27 mag.
By inspecting the HST-band flux distributions, we found \mgii\ absorbers to be, on average, more luminous than \mgii\ emitters. 
One reason for this could be that the ability to detect absorption lines depends on the strength of the continuum. 
A discussion on how this could bias our results can be found in Sect. \ref{sec:mass_sfr}.

 \cite{VanderWel2012} performed S\'ersic model fits to galaxy images selected from the CANDELS HST Multi-Cycle Treasury 
 program with the \galfit\footnote{\galfit\ homepage: \href{https://users.obs.carnegiescience.edu/peng/work/galfit/galfit.html}{https://users.obs.carnegiescience.edu/peng/work/
 galfit/galfit.html}} \citep{Peng2010} algorithm in the available near-infrared filters (H-F160W, J-F125W and Y-F105W).
We performed a positional cross-matching (within a 1 arcsec radius) between the \mgii\ sample and the \cite{VanderWel2012} catalogue.
We found that measurements of the global structural parameters were available for the majority (369/381) of our sample. 
We considered only \galfit\ fits with good quality flag \citep[good fit has quality flag equal to 0 as explained in section 4.3 of][]{VanderWel2012}.
We made use of the measurements computed in the Y-band, 
as it covers the optical rest-frame wavelength regime for our sample and typically presents higher S/N.
As already mentioned in Sect. \ref{sec:intro} we focus our discussion on the properties of \mgii\ emitters and absorbers.
We found that the two types of sources do not strongly differ in terms of b/a axis ratio, S\'ersic index and position angle, with p-values 
from a two sample KS test of 0.16, 0.65 and 0.73, respectively.
In contrast we found \mgii\ emitters to have smaller intrinsic sizes than \mgii\ absorbers, with a median value of the half-light radius 
of 1.49  compared to 3.95 kpc of the absorbers and p-value, from a two sample KS, lower than $10^{-4}$ suggesting that the two types of galaxies have 
different size distributions \citep[see also][]{Finley2017}.
It is worth noting that this is not ascribable to a difference in mean redshift (1.26 and 1.24 for \mgii\ absorbers and emitters, respectively). 
The difference in sizes suggests that there may be some physical properties that differ between the \mgii\ absorbers and emitters. We will discuss this further in Sect. \ref{sec:fitting} and \ref{sec:discussion}. 

\subsection{Emission lines detected in MUSE spectra}\label{sec:em_lines}

The emission lines detected in the MUSE spectra contain valuable information about the
physical conditions of the excited gas. The \mgii\ emitters show, on average, a well-centred (i.e. consistent with the systemic redshift) \mgii\ doublet in emission that 
can, in principle, be associated with purely nebular emission.
However, out of the \mgii\ emitters, 9 sources show an emission doublet which is redshifted from the systemic redshift of more than 50 km/s
(note that the accuracy in the velocity estimate is $\approx$40 km/s, section 4 of \citealt{Inami2017}).

As already mentioned in Sect. \ref{sec:catalogue}, the line intensities in the DR1 catalogue have been computed by assuming 
the same velocity shift relative to the systemic redshift. 
We found this setup to underestimate the \mgii\ line fluxes of \mgii\ emitters on average by 12\%. 
We recomputed these quantities for 
our \mgii\ emitters by fitting the \mgii\ line with \platefit\ accounting for the potential velocity shift.
These values have been recomputed only for the \mgii\ emitters for comparison purposes with
theoretical predictions from photoionization models (Sect. \ref{sec:nebular}). We do not study in 
this work the \mgii\ P-Cygni profile which would require a more complex fit than the 
gaussian profile used in \platefit.

In addition to the \mgii\ feature, the galaxies in the our parent sample show, depending on their redshift, collisionally excited lines, 
such as \oii, \neiii\ and \ciii, 
as well as some Balmer lines H$\beta \, \lambda 4861$, H$\gamma \, \lambda 4340$ and H$\delta \, \lambda 4101$ (hereafter H$\beta$, H$\gamma$, H$\delta$). 
Table \ref{table2} summarizes the number of galaxies where these emission lines have been detected with a line ${\mbox S/N} > 3$, from \platefit.

\begin{table*}
\caption{Additional emission lines detected in the \mgiisample}              % title of Table
\label{table2}      % is used to refer this table in the text
\centering                                      % used for centering table
\begin{tabular}{ c | c | c | c | c | c }          % centered columns (4 columns)
 & \multicolumn{4}{|c|}{\mgii\ class}  & \\
\hline\hline                        % inserts double horizontal lines
 & emitters & P-cygni & absorbers  & no detection & galaxies with \\

 & (66) & (19) & (42) & (265) & MUSE coverage\\
\hline\hline
 \oiiibpt & 8 & - & 6 & 48 & 67\\      % inserting body of the table
    \hline                                    %inserts single line
  \oiiiopt  & 7 & - & 6 &  35 & 67\\      % inserting body of the table
    \hline                                    %inserts single line
\neiiimuse & 37 & 13 & 21 & 88 & 301\\      % inserting body of the table
    \hline                                    %inserts single line
   $[\ion{O}{ii}]\lambda3726$ & 50 & 14 & 29 & 188 & 323 \\  
   $[\ion{O}{ii}]\lambda3729$ & 50 & 14 & 29 & 213 & \\
   \hline                         %inserts single line
   [\ion{C}{iii}]$\lambda1907$ & 10 & 2 & 2 & 13 & 57\\  
   \ion{C}{iii}]$\lambda1909$ & 9 & 1 & 1 & 15 & \\
    \hline
    H$\beta$ & 10 & 1 & 7 &  43 & 77\\  
    \hline
    H$\gamma$ & 30 & 8 & 24 & 95 & 215\\  
    \hline
   H$\delta$ & 30 & 10 & 26 &  71 & 239\\  
    \hline
  \end{tabular}
 \tablefoot{Number of galaxies with an emission line ${\mbox S/N} > 3$, from \platefit. The last column report the total number of galaxies of the \mgiisample\ with MUSE spectrum covering the rest-frame wavelength of the line.}             

\end{table*}
 
The most frequent other emission lines detected in the MUSE spectra of our sample are \neiii\ and \oii\ at $z\leq1.4$ and 1.5, respectively, and \ciiid\ at higher redshifts ($z\gtrsim1.44$). 
The lack of a significant number of \mgii\ emitters and absorbers with both \oiiibpt\ (hereafter \oiii) and H$\beta$ ($<10$) prevents us from exploiting their ratio,
which is sensitive to both the ionization parameter and the hardness of the ionizing spectrum, to study the excitation properties of our galaxies. 
We note that the observations of the UDF from the 3D-HST program \citep{Brammer2012, Momcheva2016} probe \oiiibpt\ for our $z>1.1$ sources.
We focus in this work on the exploitation of the MUSE spectral information and we leave the combination of MUSE-UDF and 
3D-HST grism spectroscopy to future works.

The low fraction of sources with at least two Balmer lines in their spectra makes it difficult to use their ratios to compare dust attenuation in \mgii\ emitters and absorbers.
Indeed, H$\beta$ and H$\gamma$ lines are detected with ${\mbox S/N} > 3$ in the same spectra only for 10 and 7 \mgii\ emitters and absorbers, respectively. 
Moreover, out of the 26/24 \mgii\ emitters/absorbers with both detections of H$\gamma$ and H$\delta$, 15/12 (i.e. $\sim 50\%$) have a ratio H$\gamma$/H$\delta$ lower 
than the caseB hydrogen recombination values \citep[i.e. 1.82, for an electronic temperature and density of $T=10000\,K$ and $n_{\rm  e} = 10^3$ cm$^{-3}$, from][]{Hummer1987}.
This also prevents a reliable estimate of the dust-correction from these higher order Balmer lines.
The next section will therefore focus on a deeper exploration of the \oii, \neiii\ and \ciii\ spectral features 
observed in the spectra of our galaxies.

%--------------------------------------------------------------------------------------------------------------------------
\section{Nebular emission features in \mgii\ emitters}\label{sec:nebular}

In this section, we focus on exploring the main emission features measured from the MUSE spectra of the sources classified as \mgii\ emitters 
to verify whether their emission is consistent with ionization by photons produced in \hii\ regions. 
We rely on predictions from photoionization models of star-forming galaxies, described in Sect. \ref{sec:models}. 
In the following subsections we then show how these calculations compare with the observed emission line 
ratios and, for completeness, inspect model predictions of ionizing sources of different origins, such as AGN and radiative shocks. 

\subsection{Photoionization models of star-forming galaxies}\label{sec:models}

\cite{Gutkin2016} recently built a comprehensive set of synthetic models of stellar and nebular emission from a whole galaxy by 
 combining the spectral evolution of typical, ionization-bounded, \hii\ regions with a star formation history.
 
To obtain the emission of different \hii\ regions, powered by newly-born star clusters, 
these calculations combine the latest version of the 
stellar population evolutionary synthesis models of \cite{Bruzual&Charlot2003}, 
Charlot \& Bruzual, in prep., with the photoionization code CLOUDY c13.03 \citep{Ferland2013}, following the approach first outlined in \cite{Charlot&Longhetti2001}.
The new update to the \cite{Bruzual&Charlot2003} synthesis models incorporates new stellar evolutionary tracks from \cite{Bressan2012}, 
including the evolution of massive Wolf-Rayet stars and new stellar spectral libraries 
(see section 2.1 of \citealt{Gutkin2016} for more details on the stellar emission and 
\citealt{Wofford2016} for a comparison with other spectral synthesis models).

The \cite{Gutkin2016} models relate the gas-phase metallicity measured from nebular emission lines to the total 
(both gas- and dust-phase) interstellar metallicity of the ionized medium
through a self-consistent treatment of element abundances and depletion onto dust grains. 
The interstellar abundances and depletion factors are listed in Table 1 of \cite{Gutkin2016}. 
The models are parametrized in terms of the following physical quantities (see also Table 3 of \citealt{Gutkin2016} for a summary of the full-grid parametric sampling): 

\begin{itemize} 
\item the volume-averaged ionization parameter, \logU, defined as the dimensionless ratio of the number density of
H-ionizing photons to that of hydrogen, ranges between -3.65 and -0.65 in logarithmically spaced bins of 0.5 dex;
\item the hydrogen gas density, $n_{\rm H} =10, 10^2, 10^3$ and $10^4$ cm$^{-3}$;
\item the interstellar (i.e. gas+dust-phase) metallicity $Z$ (assumed to be the same as the stellar component), ranges from 0.0001 to 0.04 (the total present-day solar metallicity adopted is $Z_{\odot} = 0.01524$);
\item the dust-to-metal mass ratio, $\xi_{\rm d}= 0.1, 0.3$ and $0.5$, sets the fraction of heavy elements depleted onto
dust grains;
\item the carbon-to-oxygen ratio, C/O, from 0.1 to 1.4 times the solar values (C/O)$_{\odot} =0.44$;
\item the upper mass cut-off, $M_{\rm up} = 100$ and 300 M$_{\odot}$, of the initial mass function (IMF), assumed to be a Galactic-disc IMF from \cite{Chabrier2003}.

\end{itemize} 
  
We note that, following the definition in Eq. B.6 of \cite{Panuzzo2003}, the volume-averaged ionization parameter is a factor of 9/4 larger than 
the ionization parameter $U_{\rm S}$ listed in Table 3 of \citet[][see footnote of \citealt{Hirschmann2017}]{Gutkin2016}.
In the next subsection we compare a sub-grid of these models with the spectral measurements of \mgii\ emitters.   
These models have also recently been incorporated in the spectro-photometric fitting tool \beagle\ (BayEsian Analysis of GaLaxy sEds), 
whose main features are summarized in Sect. \ref{sec:beagle}.

\subsection{Comparison with observations in \mgii\ emitters}\label{sec:emitters}

%-----------------------------
% Figure 3
  \begin{figure*}
  \centering
  \includegraphics[width=14cm]{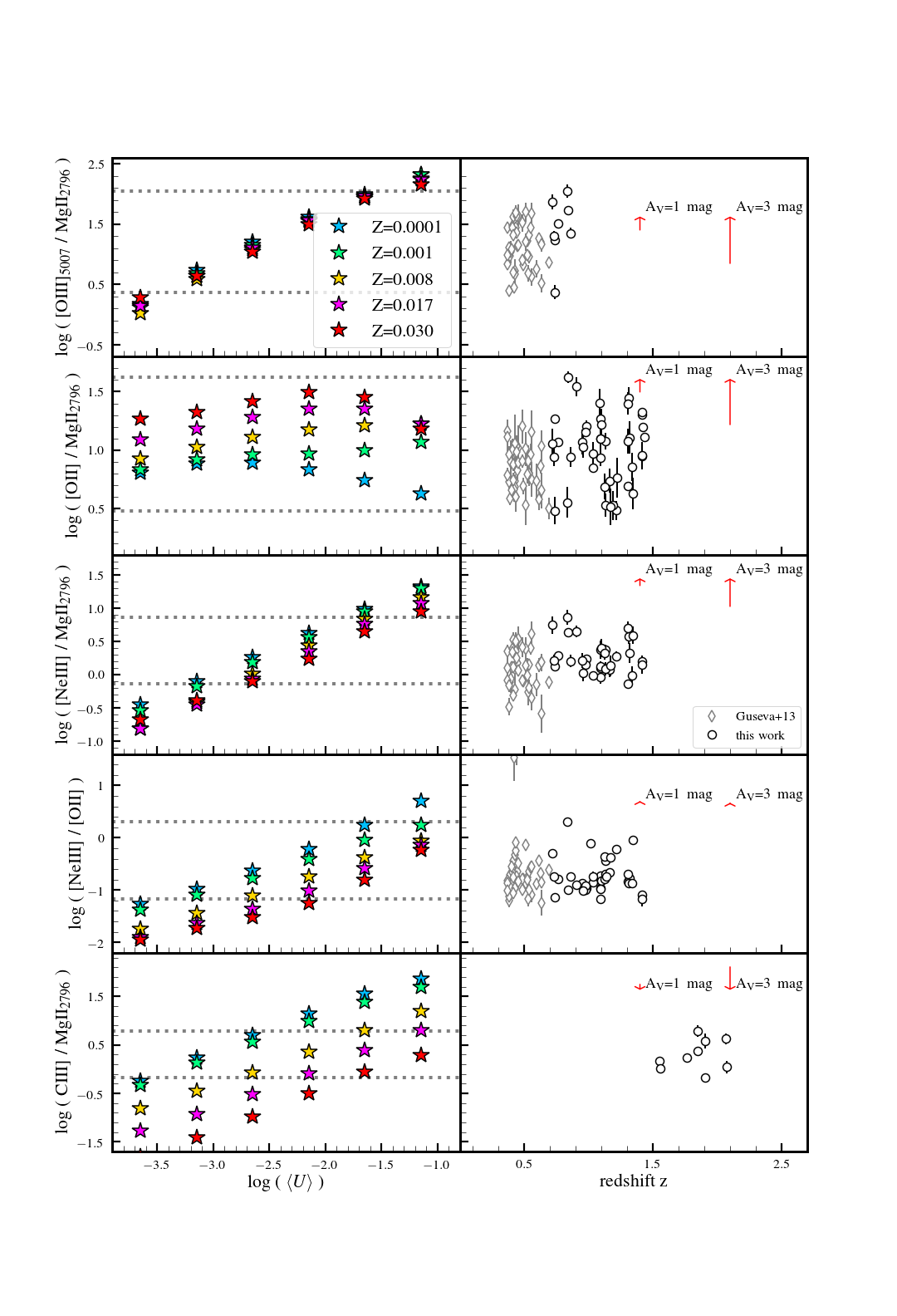}
   \caption{Left:  \oiiibpt/\mgiib, \oii/\mgiib, \neiii/\mgiib, \neiii/\oii\ and \ciii/\mgiib\ emission-line ratios predicted 
   from the star-forming galaxy models of \cite{Gutkin2016}, described in Sect. \ref{sec:models},  
   for different values of the volume-averaged ionization parameter \logU\ (x-axis) 
   and metallicity Z (colour coded as indicated in the top panel). Gray dotted lines mark the 
   minimum and maximum value of the line ratios measured from the MUSE spectra of \mgii\ emitters.
 Right (from top to bottom): observed line ratios 
 as function of redshift for the \mgii\ emitters, as defined in Sect. \ref{sec:classification} (empty-black circles) 
 and for the sample of \cite{Guseva2013} (empty-gray diamonds). 
Red arrows in the top-right of each panel indicate the effect of attenuation by dust for A$_{\mbox v} = 1$ and $3$ mag and 
 a \cite{Calzetti2000} attenuation curve. Data measurements from \cite{Guseva2013} are corrected for dust attenuation, while the MUSE fluxes are not.}
            \label{Fig3}%
   \end{figure*}

The five panels of Fig. \ref{Fig3} show how predictions of the \cite{Gutkin2016} models (left panels)
compare with the emission-line ratios measured from the MUSE spectra of our \mgii\ emitters (right panels), 
namely, from top to bottom, \oiiibpt/\mgiib\ (mainly for comparison purposes with low redshift samples), \oii/\mgiib, \neiii/\mgiib, \neiii/\oii\ and \ciii/\mgiib. 
Note that, for our comparison, we considered here the blue component of the \mgii\ doublet, \mgiib, as the red component \mgiir\ is detected with 
${\mbox S/N} > 3$ only on 31/63 \mgii\ emitters. Indeed its theoretical value is, depending on the optical depth,
one to two times the intensity of the red component, \mgiir\ \citep[e.g.]{Laor1997}.
The median ratio of the \mgii\ emitters with both components of the doublet detected with ${\mbox S/N} > 3$ is 1.64.

The right and left panels of Fig. \ref{Fig3} show data measurements and model predictions, respectively.
Data measurements comprise both our MUSE \mgii\ emitters (black empty circles) 
and lower redshift $0.36<z<0.7$ galaxies (gray empty diamonds) from the sample of \cite{Guseva2013},
described in Sect. \ref{sec:intro} and \ref{sec:classification}.
As the line ratios of MUSE \mgii\ emitters are not corrected for dust attenuation,
the red arrows in each right panel show the effect of dust reddening for attenuation in the V-band of $A_{\rm V}=1$ and $3$ mag and
a \cite{Calzetti2000} attenuation curve.

The models of \cite{Gutkin2016} are shown for dust-to-metal mass ratio $\xi_{\rm d}=0.3$ \citep[average value of the][model grid]{Gutkin2016}, 
hydrogen density $n_{\rm H}=10^{2}$ cm$^{-3}$, solar $C/O$ ratio, upper mass cut-off $M_{\rm up}=100M_{\odot}$ and
for a variety of volume-averaged ionization parameters (x-axis of left panels)
and metallicities $Z$ (color-coded as labeled in the third panel).
The gray dashed lines indicate the minimum and maximum value of the ratios observed in our \mgii\ emitters.
We found that the set of models predict line ratios similar to the observed ones.

We do not aim here at an in-depth comparison between the ratios measured for our sources and those at lower redshift, 
but we note that the observed \neiii/\oii\ ratios of our \mgii\ emitters are similar, within the measurements uncertainties,
to those of lower redshift galaxies from \cite{Guseva2013}.
On the contrary, the mean  \oiii/\mgiib\ and \neiii/\mgiib\ ratios appear to be larger for the MUSE \mgii\ emitters. 

At the same time, we also note that data measurements, including those from \cite{Guseva2013}, are not corrected for absorption from the ISM which could
strongly affect the intensity of the \mgii\ doublet \citep[see e.g. Table 1 of ][]{Vidal2017}, leading to an underestimate of the nebular flux. 

For illustrative purposes, Fig. \ref{Fig3} shows only a sub-grid of the \cite{Gutkin2016} models. It is worth noting, however, that our full suite of models
allows for a better coverage of the parameter space. 
We did not find the ratios considered here to strongly depend on variations of hydrogen gas density and upper mass cut-off. 
Also, no dependence has been found for the C/O ratio. 
The latter is because we do not probe Oxygen and Carbon lines in the same spectrum.
These three parameters are kept fixed while performing the fits to the observed line fluxes (see Sect. \ref{sec:fitting}).
On the contrary, \mgii\ is a refractory element and, hence, is sensitive to metal depletion onto dust grains. 
This is the reason why we let the dust-to-metal mass ratio freely vary in the spectral fitting (Sect. \ref{sec:fitting}). 

The \oii/\mgiib\ ratio is shown to be more sensitive to metallicity (second left panel of Fig. \ref{Fig3}) than other ratios and, more importantly, 
more sensitive to metallicity than to other parameters (in particular $\langle U \rangle$).
The rise of the \oii/\mgiib\ ratio as metallicity increases follows from the increase in the abundance of coolants (such as Oxygen). 
Towards high metallicities, the increase of Oxygen abundance is compensated by an higher efficiency of cooling. 
This makes the electronic temperature to drop. 
As consequence, the \mgii\ emission compared to that of \oii\ is reduced because \mgii\ requires a higher potential (~4.4 eV) for collisional excitation than \oii\ (~3.3eV).
The wide range of \oii/\mgiib\ values of our \mgii\ emitters, ($0.5 < $log (\oii/\mgiib)$< 1.6$),
suggests a variety of metallicities within our sources.
Similarly, the spread of the model points shows that the \ciii/\mgiib\ ratio is also sensitive to metallicity.
The stellar ionizing spectra is harder at lower metallicity, producing enough high energy photons to ionize and excite \ciii\ and, hence, increase the \ciii/\mgiib.

Combining these ratios with the other ones more sensitive to other physical quantities, such as the ionization parameter, 
will provide useful constraints on the physical properties of the ionized gas. 
For example, \oiii/\mgiib\ and \neiii/\mgiib, which show little dependence on metallicity, provides important constraints 
on the ionization parameters of our galaxies (top and middle panels of Fig. \ref{Fig3}).
The \neiii/\oii\ ratio has been used both as metallicity \citep[e.g.][]{Nagao2006, Maiolino2008} and ionization potential indicator \citep[e.g.][]{Ali1991, Levesque2014} indicator.
We note that the use of an emission-line ratio as probe of a physical quantity depends on the models assumptions \citep{Levesque2014}.
For the models considered in this work, the \neiii/\oii\ ratio can provide a certain level of constraint in the ionization paramater but it is also degenerate with metallicity.
The observed data are compatible with a volume-averaged ionization parameters 
ranging from \logU$\, \sim -3.2$ up to -1.5, i.e. spanning values comparable to those of star-forming
and intensively star-forming galaxies \citep[e.g.][]{Stasinska1996,Brinchmann2004,Brinchmann2008,Shirazi2012}, 
and up to higher values commonly 
observed in young compact star-forming galaxies \citep[e.g.][]{Stark2014,Izotov2016,Izotov2017,Chevallard2017}.
In conclusion, these ratios give no indication that we need to invoke harder ionization sources than massive stars, such as AGN or shocks,
to reproduce the observed ratios of our MUSE \mgii\ emitters. 
Nevertheless, for completeness, we now explore other types of ionizing sources.

\subsection{Contribution from other ionizing sources}\label{sec:others}

In addition to the previous subsection, 
we have compared the observed \oii/\mgiib, \neiii/\mgiib, \neiii/\oii\ and \ciii/\mgiib\ ratios with 
predictions from photoionization models of narrow-line emitting regions in AGN \citep{Feltre2016} and of shocks \citep{Allen2008}.
We found that these models can predict the emission-line ratios observed in the spectra of our galaxies and none of these ratios 
enable a proper distinction between the different types of ionizing source.
Complementary information is required to better explore any contribution from other ionizing sources, 
such as additional optical and UV emission-lines that would provide more constraints on the excitation properties of these sources. 

As mentioned in Sect. \ref{sec:sample}, to quantify the AGN contamination, we positionally cross-matched the whole \mgiisample\ 
with the 7 Ms Source Catalogs of the Chandra Deep Field-South Survey \citep{Luo2017}. We found an X-ray counterpart only for one \mgii\ emitter, 
ID872, which has been discarded from our sample as it exhibits broad \mgii\ emission (see Sect. \ref{sec:sample}).
Hence, we exclude nuclear gravitational accretion as the dominant source of ionization in our \mgii\ emitters, although the presence of low-luminosity 
or heavily obscured AGN might not be completely excluded \citep{Luo2017}. 

Unfortunately, we do not possess enough information to rule out a potential contribution from radiative shocks  
to the spectra of \mgii\ emitters. 
In the literature, \mgii\ emission accompanied by blueshifted absorption, in a P-Cygni like profile, has been commonly used as tracer of galactic winds \citep[e.g.][]{Weiner2009,Rubin2010,Rubin2011,Erb2012,Finley2017}.
These sources are not included in the \mgii\ emitters, but they are considered here separately, as explained in Sect. \ref{sec:classification}.
In addition, shocks are usually associated with intense star formation or AGN and are generally not expected to be the dominant source of line emission within a galaxy \citep[e.g.][]{Kewley2013}. 
We therefore conclude that, even if radiative shocks could still be present in our \mgii\ emitters, their contribution to the typical total spectrum is unlikely to be dominant.
In this respect, spatially resolved spectroscopy has been proven to be extremely useful to study the impact of shock contamination on the emission lines measured from galaxy spectra \citep[e.g.][]{Rich2011,Rich2014,Yuan2012}.

%---------------------------------------------------------------------------------------------------------------------------------
\section{Spectral Fitting Analysis and Results}\label{sec:fitting}

Emission-line ratios contain valuable clues on the source of ionization and the physical properties of the ionized gas. 
However, to probe other physical quantities of the galaxies, including stellar mass, SFR and dust attenuation, one needs to 
combine, in a self-consistent way, both continuum and nebular emission from stars and gas.
With the aim of inferring these properties, we relied on SED fitting technique. 
Specifically, in this work we used the fitting code \beagle\ \citep{Chevallard2016} which already incorporates the models of \cite{Gutkin2016}
described in Sect. \ref{sec:models}.

In what follows, we first provide a brief overview of the main features of the \beagle\ tool, along with the input settings chosen for the purposes of this work,
and then present the main results from this spectral analysis.

\subsection{Spectrophotometric fitting tool \beagle}\label{sec:beagle}

\beagle\ \citep{Chevallard2016} is a flexible tool, built on a 
Bayesian framework, to model and interpret the SED of galaxies.
Briefly, the current version of this code self-consistenly incorporates the continuum radiation emitted by stars 
within galaxies and the reprocessed nebular emission from the ionized gas in \hii\ regions.
It also accounts for the attenuation by dust and the transfer of radiation through the intergalactic media 
and includes different prescriptions to treat the chemical enrichment and star formation histories of galaxies.
\beagle\ can be exploited to interpret any combination of photometric and spectroscopic data from the UV to the near-infrared range, 
as well as to build mock spectra of galaxies (see also Sec. \ref{sec:disc_models}).

\subsection{\beagle\ fitting to the \mgiisample}

In this section we explore the synergy of HST and MUSE by
simultaneously fitting broad-band photometry and integrated fluxes for our \mgiisample, 
described in Sect. \ref{sec:sample}. 
The 11 bands from HST allow us to constrain the stellar and recombination continuum, while
the spectral information from MUSE provide useful information on the nebular emission. 
By applying the \beagle\ tool to our \mgiisample, we aim at deriving the properties of our galaxies
(such as stellar mass, SFR, dust attenuation), and explore to which extent we can probe 
the physical properties of the ionized gas (such as metallicity and ionization parameter) 
with our observations. 
In the following, the values reported for each physical properties inferred from the fit correspond to the posterior
median and the errors indicate the 68\% central credible interval.

\subsection*{Spectro-photometric data}

We simultaneously fit the HST broad-band photometry (see Sect. \ref{sec:hst})
and the integrated fluxes measured from the MUSE spectra for the whole \mgii\ sample.
We considered the strongest emission lines detected, with a ${\mbox S/N} > 3$ from \platefit, in our spectra (see Table \ref{table2} and Sect. \ref{sec:em_lines}).
Since a modelling of the neutral ISM needed for the treatment of resonant lines like \mgii\ is not yet incorporated in \beagle\ we do not include the \mgii\ doublet in the fitting.

In the fitting procedure, the observed line intensities are compared with the integrated line fluxes computed on the spectral models (lines+continuum) incorporated within \beagle.
These models also include the stellar features which are, instead, already 
subtracted from line fluxes computed with \platefit. We therefore can not directly input the \platefit\ line fluxes into \beagle.
We computed the integrated fluxes of the lines detected on the MUSE spectra (Table \ref{table2})
using the {\it MUSE Python Data Analysis Framework} \footnote{\url{https://git-cral.univ-lyon1.fr/MUSE/mpdaf}} (MPDAF).
Specifically, we performed a Gaussian fit to the section of the spectrum that contains the expected emission line using 
the \texttt{gauss$\_$fit} (and  \texttt{gauss$\_$dfit} for line doublets) function of MPDAF.

%-----------------------------
% Figure 4
  \begin{figure*}
  \centering
        \includegraphics[width=18cm]{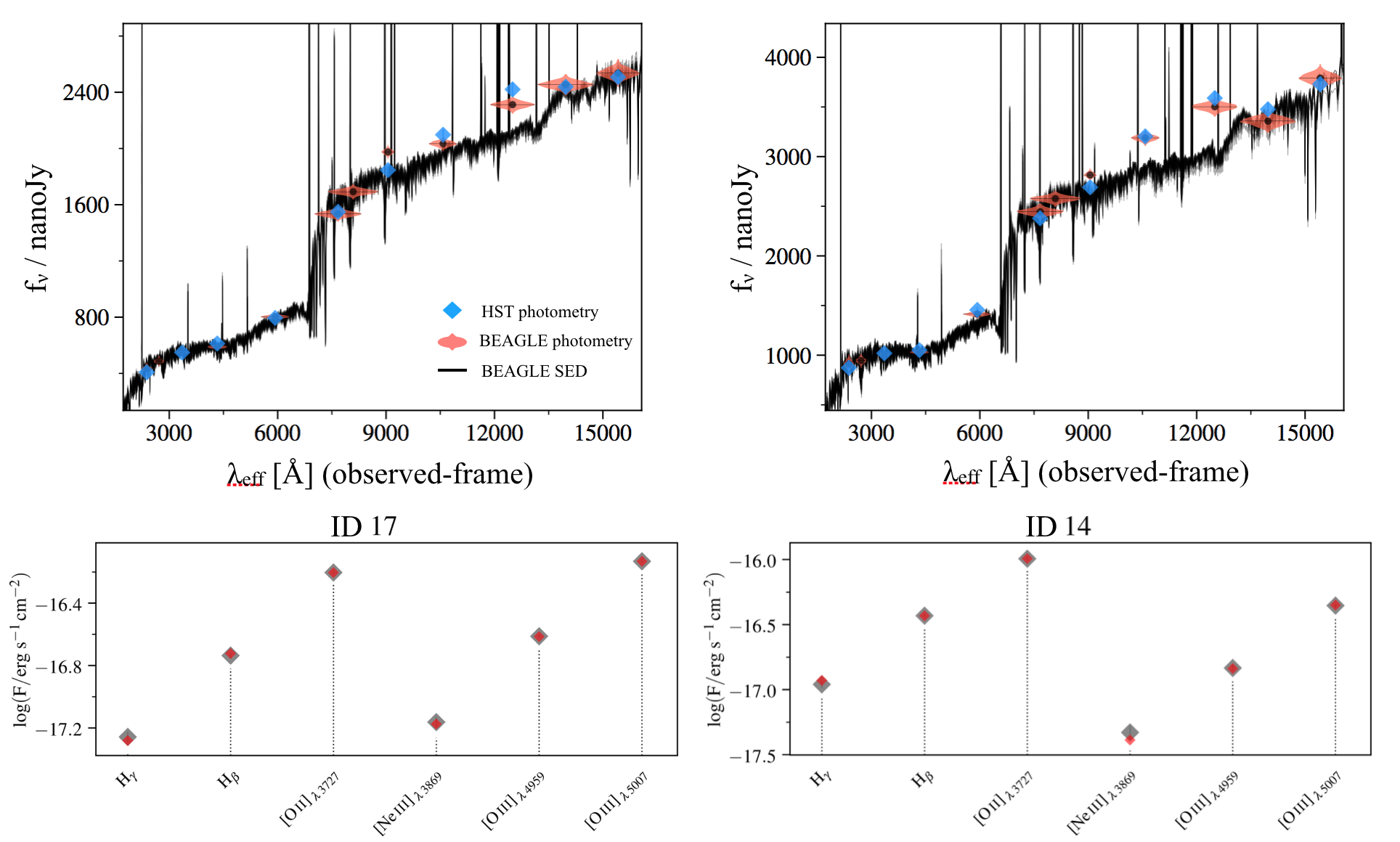}
   \caption{Example of a simultaneous \beagle\ fit to HST photometry (top) and MUSE integrated fluxes (bottom) for the \mgii\ emitter ID17 at z$=0.84$ (left) and \mgii\ absorbers ID14 at z$=0.77$ (right). Top: HST (cyan diamonds) and predicted (black points and red shaded area) broad-band photometry. In black the full SED predicted from the \beagle\ fit. Bottom: integrated fluxes measured from MUSE spectra (red diamonds) and from the SED predicted by \beagle\ (gray diamonds). Error bars on the data points are contained within the markers.}
            \label{Fig4}%
   \end{figure*}

\subsection*{\beagle\ settings}

We adopted stellar models (Sect. \ref{sec:models}) computed using a standard \cite{Chabrier2003} IMF with 100 \msun\ as upper mass cut-off.

We kept the hydrogen gas density of the clouds fixed at $n_{\rm H}=10^2\, {\rm cm^{-3}}$ and the C/O abundance ratio at the solar value 
for the reasons explained in Sect. \ref{sec:nebular}.
We assumed a delayed star formation history $\psi (t) \propto t\, {\rm exp}(-t/\tau_{\rm SFR})$, for any age $t$ 
over the galaxy lifetime \citep[section 4.2 of][]{Chevallard2016}, where $\tau_{\rm SFR}$ is the star formation time scale.

We followed the model of \cite{Charlot2000} to describe dust attenuation. This models assumes two components, 
one associated with the short-lived birth clouds and another one diffuse throughout the ISM. Dust attenuation is parametrized
in terms of the total optical depth, $\tau_{\rm V}$, and the fraction of attenuation, $\mu$, due to the diffuse ISM \citep[see figure 5 of ][to see how these
absorption curves impact the computation of the UV spectral slope $\beta$, reported in Sec. \ref{sec:results}]{Charlot2000}.
We also explored a different approach for the dust attenuation, i.e. the `quasi-universal' relation of \cite{Chevallard2013}
between the shape of the attenuation curve and the V-band attenuation optical depth in the diffuse ISM, which accounts for geometrical effects and galaxy inclination.
However, we found no difference related to the dust prescriptions in the qualitative trends discussed in this section.

We let the following several adjustable physical quantities of the models incorporated in \beagle\ vary freely, assuming uniform prior distributions
in either logarithmic or linear quantities, as indicated below:

\begin{itemize}
\item interstellar metallicity Z ($-2.2 \leq {\rm log }(Z/$\zsun$) \leq 0.24$);
\item volume-averaged ionization parameter ( $-3.65 \leq $\,\logU\,$\leq -0.65$);
\item the dust-to-metal mass ratio ($0.1 \leq \xi_{\rm d} \leq 0.5$);
\item the star formation timescale, $\tau_{\rm SFR}$, from 7 to 11.5 Gyr;
\item  V-band dust optical depths in the range $ -3. \leq {\rm log}\, \tau_{\rm V} \leq 0.7$;
\item the fraction $\mu$ of attenuation arising in the diffuse ISM \citep{Charlot2000}, which ranges from 0 to 1.
\end{itemize}

\subsection{\beagle\ Results} \label{sec:results}

Fig. \ref{Fig4} shows two examples of a simultaneous fit to HST broad-band photometry (left panel) and MUSE integrated fluxes (right panel) for
the \mgii\ emitter ID17 ($z=0.84$) and the \mgii\ absorber ID14 ($z=0.77$).
We inferred several galaxy properties from the spectral fitting, as discussed below, such as stellar mass, 
star formation rate (SFR, averaged over the last $100$ Myr) and specific star formation rate (sSFR), dust optical depth and ionizing emissivity.

We focused our analysis on the comparison of the properties of \mgii\ emitters with those of absorbers, 
leaving a detailed study of the \mgii\ P-Cygni to future works (Finley et al., in prep). 
We typically found the volume-averaged ionization parameter to be $-3.4 \lesssim$\,\logU\,$\lesssim -2.0$
and the metallicity $0.1 \lesssim Z/Z_{\odot} \lesssim 1.5$, in agreement with Fig. \ref{Fig3}.
We found \mgii\ emitters and absorbers to have similar distributions of the gas nebular properties, namely metallicity, 
ionization parameter and dust-to-metal mass ratio. This favours a scenario in which \mgii\ emission is a tracer of specific galaxy properties not necessarely 
connected to the properties of the gas within the ionization regions. 
In the following sections, we discuss the most interesting results from our spectral fitting analysis. 

\subsection*{Stellar Mass and Star Formation}\label{sec:mass_sfr}

%-----------------------------
% Figure 5
  \begin{figure}
   \hspace*{-0.7cm}    
   \includegraphics[width=10.5cm]{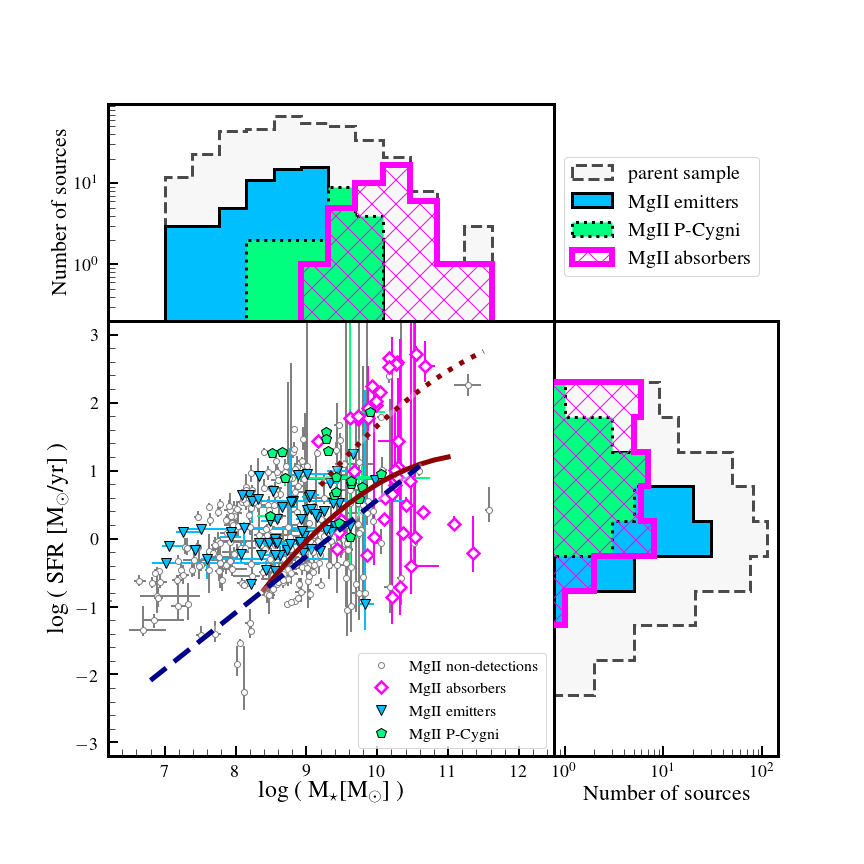}
   \caption{Main sequence, SFR versus stellar mass, for the \mgiisample. We observe a smooth transitions in stellar mass from \mgii\ emitters (cyan filled triangles),  to P-cygni (green filled pentagons), to absorbers (magenta empty diamonds). Gray circles are \mgii\ non-detections. Also shown main-sequence curves from \cite{Boogaard2018}, submitted for z$\sim$0.9 (dashed dark blue line), and \cite{Whitaker2014} for $0.5 < z < 1.0$ and $1.5 < z < 2.0$ (solid and dotted dark red curves, respectively). }
            \label{Fig5}%
   \end{figure}

Fig. \ref{Fig5} shows the stellar mass versus SFR sequence for the \mgiisample. 
The typical (median) errors on stellar mass and SFR ranges from $\sim \pm 7\%$ to $\sim \pm 15\%$, and from $\sim \pm 12\%$ to $\sim \pm 24\%$, respectively. 
For reference we report the star formation sequence by \cite{Whitaker2014} for $0.5 < z < 1.0$ and $1.5 < z < 2.0$ (solid and dotted dark red curves, respectively) derived from a mass-complete sample of star-forming galaxies
in the CANDELS fields, drawn from the 3D-HST photometric catalogues. Specifically, we used the polynomial fit coefficients that parametrize the evolution of the SFR-stellar mass sequence reported in Table 1 of \cite{Whitaker2014}. 
We also show the star formation sequence at $z\sim0.9$ from equation 11 of Boogaard et al. 2018, submitted (dark blue dashed line) computed using MUSE observations of the {\it Hubble} Ultra Deep Field 
and the {\it Hubble} Deep Field South of a sample of galaxies at $0.11 < z < 0.91$, with stellar masses between $10^7$ and $10^{10.5} M_{\odot}$.

Similarly to previous findings, we found \mgii\ emitters to exhibit, on average, lower stellar masses than the absorbers (see histograms in Fig. \ref{Fig5}).
The median values of stellar mass are $5.9 \times 10^8 M_{\odot}$ and $1.6 \times 10^{10} M_{\odot}$ for \mgii\ emitters and absorbers, respectively. 
A two sample KS test gives a p-value lower than $10^{-18}$, so we choose to reject the null hypothesis that the two samples have the same stellar mass distributions.
Analogous results were also found by previous works: \cite{Erb2012} on a sample drawn from a survey carried out with the LRIS spectrograph (LRIS-B) on the Keck I Telescope \citep{Steidel2004}, 
\cite{Kornei2013} in star-forming galaxies at $z\sim1$ from the DEEP2 survery and by \cite{Finley2017} in a subsample of this \mgiisample.
Here we further confirm these previous findings by merely selecting our sources on the \mgii\ line, without any previous selection
on photometric colours or other emission lines. 
We note that MUSE has allowed us to probe \mgii\ emission in galaxies with stellar masses one up to two orders of magnitude lower than those explored in the previous studies of \cite{Erb2012} and \cite{Kornei2013}.

The \mgii\ absorbers also reach higher values of SFR, and lower values of sSFR, compared to emitters.
Fig. \ref{Fig5} is in overall agreement with Fig. 3 of \cite{Finley2017} where stellar masses and SFR were obtained with two different methods (spectral fitting and
an empirical relation using \oii-dust corrected).
It is worth noting that despite some quantitative differences between the SFR inferred from \beagle\ and those obtained with other methods, 
the general trends remain unchanged. 
Interestingly, galaxies with \mgii\ P-Cygni profile show intermediate properties between emitters and absorbers and will be subject of future studies. 

Before going ahead with the interpretation of the results, we need to consider the potential biases
introduced by our sample selection. Indeed, \mgii\ absorbers are, on average, more luminous than \mgii\ emitters,
because the ability to detect absorption features depends on the strength of the continuum. 
Moreover, the detection of continuum-faint \mgii\ emitters at $z>1.5$ is complicated by skyline residuals. 
We performed a first test by dividing the sample in two redshift bins, $0.7 < z \leq 1.5$ and $1.5 < z \leq 2.34$, 
and then selecting \mgii\ emitters and absorbers within a given range of continuum luminosities in the F606W band.

\begin{figure}
  \centering
   \includegraphics[width=9.5cm]{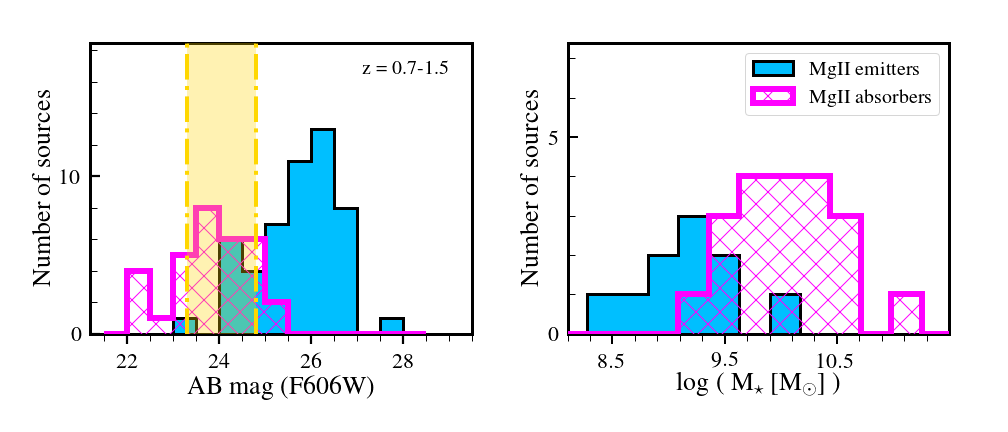}
   \caption{Left: F606W passband filter flux distribution for \mgii\ emitters and absorbers (cyan and magenta histograms, respectively) in the redshift range $0.7<z<1.5$. Right: Stellar mass distributions of \mgii\ emitters and absorbers (same color-code as the left panel) with a given range of F606W flux, as highlighted in yellow in the left panel. }
            \label{Fig6}%
   \end{figure}

The lower redshift bin, $0.7 < z < 1.5$, had more than 10 sources per \mgii\ spectral type (emitter and absorber) in the 
range of luminosity $23.3 <  {\mbox F6060W} <  24.8$ (where 23.3 is the brightest common magnitude between \mgii\ emitters and absorbers in the redshift range of interest, and 24.8 is the 
peak value of F6060W distribution of the \mgii\ absorbers), allowing for a statistical comparison. 
The F606W flux distributions of \mgii\ emitters and absorbers for the lower redshift bin are shown in Fig. \ref{Fig6} (left panel). 
As can be seen from the right panel of Fig. \ref{Fig6}, the \mgii\ emitters and absorbers within the selected range of F606W continuum fluxes 
(yellow shaded area in Fig. \ref{Fig6}, left panel) 
still show a dichotomy in stellar mass. 

The p-value from a two sample KS test for this sub-sample is lower than $2\times10^{-4}$ and allows us to
reject the null hypothesis that the two mass distributions are the same. 
We obtained the same result when considering HST F775W.
For the higher redshift bins there is, unfortunately, low number statistics. The number of galaxies with $1.5 < z \leq 2.34$ and within 
a common range of luminosity is limited to 5 absorbers and 9 emitters.
We performed anyway a two sample KS test and found a p-values lower than 0.008 both for HST F606W and HST F775W. 
This suggests that, even though our sample is likely not to be complete in terms of low-luminosities \mgii\ absorbers (see Sect. \ref{sec:sample_selection}), 
selection effects alone seem to do not be the primary driver for the difference in stellar mass between \mgii\ absorbers and emitters (Finely et al., in prep). 

\subsection*{Equivalent Width of \mgii}\label{sec:mass_sfr}

\begin{figure*} 
   \includegraphics[width=18cm]{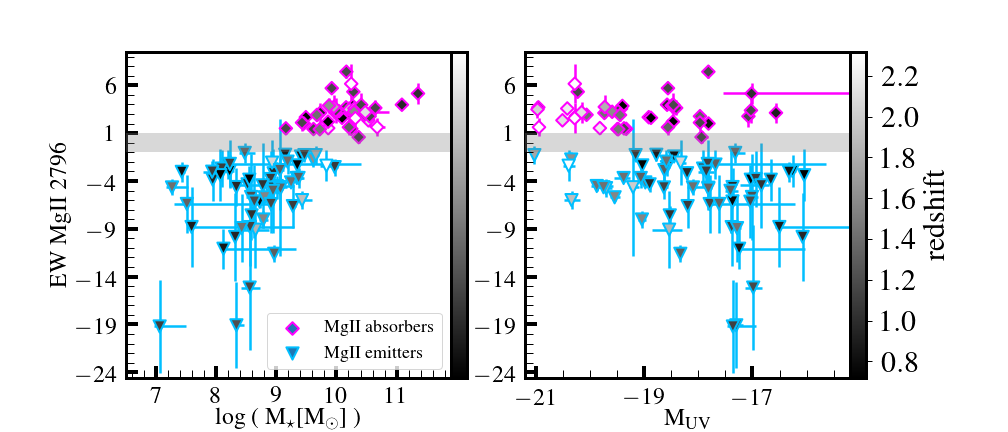}
   \caption{EW of \mgiib\ for \mgii\ emitters (cyan triangles) and absorbers (magenta diamonds) as function of the stellar mass (left) and UV absolute magnitude at 1600 $\AA$ (right), color-coded accordingly to the redshift. 
   The gray shaded area indicates the threshold values used to identify \mgii\ emitters (EW $< -1.0$) and absorbers (EW $> 1.0$).}
            \label{Fig7}%
   \end{figure*}

Fig. \ref{Fig7} shows the EW of the \mgiib\ doublet component (computed with \platefit\ as described in Sect. \ref{sec:catalogue}) versus stellar mass (left), inferred from the fit
and UV absolute magnitude at 1600 $\AA$ (right), computed on the SED predicted by \beagle, 
for \mgii\ absorbers and emitters, defined to have EW \mgii\ $>+1$ and $<-1$, respectively (see Sect. \ref{sec:classification}).
\mgii\ emitters with high masses do not have strong \mgiib\ EW in emission (left panel of Fig. \ref{Fig7}). 
This could be explained in a scenario where, as the amount of ISM increases with stellar mass, the emission diminishes until it becomes completely suppressed, 
as we discuss in Sect. \ref{sec:discussion}.
Moreover, there is a lack of strong \mgiib\ EW in emission for the bright (right panel of Fig. \ref{Fig7}) \mgii\ emitters.

\mgii\ is very sensitive to emission-infill \citep[e.g.][]{Prochaska2011, Scarlata2015, Zhu2015, Finley2017}, due to re-emission of 
photons of the same same wavelength of the transition (\mgiit\ in this case) that fills in the absorption profile. 
To correct EW measurements of \mgii\ for emission-infill, \cite{Zhu2015} proposed an observation-driven method
which consists in comparing EW of \mgii\ and Fe {\sc ii} $\lambda 2344, \lambda 2374, \lambda 2586, \lambda 2600$ detected in quasar absorption-line
systems to those observed in the spectra of star-forming galaxies. 
We do not cover the Fe {\sc ii} transitions for all the sample and we do not aim here at quantitatively discussing the EW measurements for \mgii\ absorbers.
But we note that corrections for emission-infill for the most massive sources would increase the EW of \mgii\ absorbers shown in Fig. \ref{Fig7} by values
between $1.6 - 3 \,\AA$ \citep[][Finley et al. in prep.]{Finley2017}.
These corrections will introduce additional dispersion to Fig. \ref{Fig7}, but will not impact the results discussed in Sec. \ref{sec:discussion}.

%-----------------------------
% Figure 6
%

%-----------------------------
% Figure 7

\subsection*{Dust Attenuation, UV Spectral Slope and Ionizing Emissivity}\label{sec:attenuation}

%-----------------------------
% Figure 8
  \begin{figure}
  \centering
   \includegraphics[width=9cm]{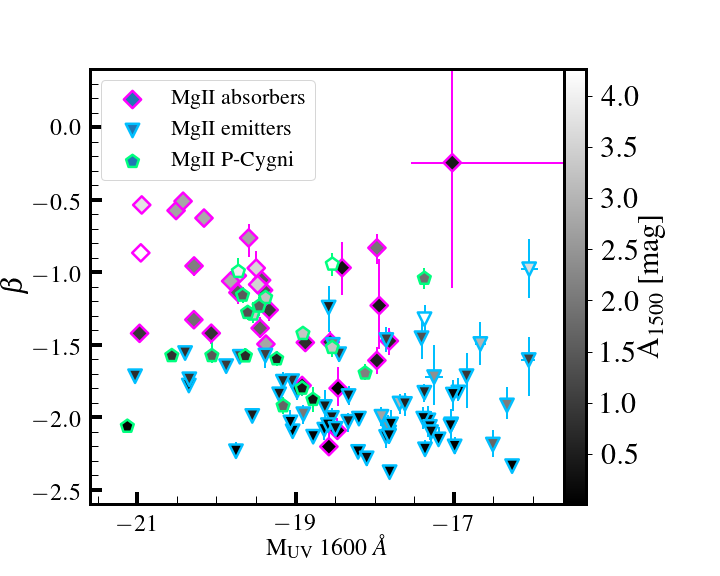}
   \caption{UV spectral slope, $\beta$, versus UV absolute magnitude at 1600 $\AA$ and color-coded as function of the attenuation at 1500 \AA, computed from the \beagle\ SED. 
   Cyan triangles, green pentagons and magenta diamonds are \mgii\ emitters, P-cygni and absorbers, respectively. }
            \label{Fig8}%
   \end{figure}

The dust attenuation at $1500\,\AA$, $A_{1500}$, inferred from the fitting, is on average higher for \mgii\ absorbers, 
with a median value of $\sim 1,52$ compared to the $\sim 0.38$ mag of the emitters.
\cite{Kornei2013} also found galaxies with strong \mgii\ emission to have lower dust attenuation than their whole sample. 

From the SED predicted from the \beagle\ fits, we computed the UV spectral slope $\beta$ (defined as $F_{\lambda} \propto \lambda^{\beta}$) following the parametrization of \cite{Calzetti1994} 
and the intrinsic (non corrected for dust attenuation) UV luminosity at 1600 $\AA$, M$_{\rm UV}$. 
As the spectral slope is particularly sensitive to the dust content within the galaxy, we expect the UV slopes to be, on average, bluer for \mgii\ emitters than absorbers.
This is shown in Fig. \ref{Fig8}, color-coded accordingly to the dust attenuation.
\mgii\ emitters have UV spectral slope $\beta < -1 $, with a median value of $\sim -1.98$, for a wide range of UV luminosities ($-21.0 < $M$_{\rm UV} < -16.6$).

The \mgii\ absorbers, which are on average UV-brighter than emitters in our sample (Sect. \ref{sec:sample}), are also the more massive (Sect. \ref{sec:hst}) and have redder UV spectral slopes with increasing UV luminosity. 
At fixed UV luminosity, \mgii\ emitters have bluer spectral slopes than absorbers. 
This is in agreement with previous findings from \cite{Erb2012}, even though their \mgii\ emitters also included sources with P-Cygni profile which do not show any particular trend 
in Fig. \ref{Fig8} but lie in between \mgii\ emitters and absorbers. 

From the \beagle\ fits, we inferred the ionizing emissivity, 
i.e. the number of ionizing photons per UV luminosity. 
We computed two different ionizing emissivities considering both the unattenuated and attenuated UV luminosity, as follows:

%-----------------------------
% Figure 9
  \begin{figure*}
  \centering
   \includegraphics[width=15cm]{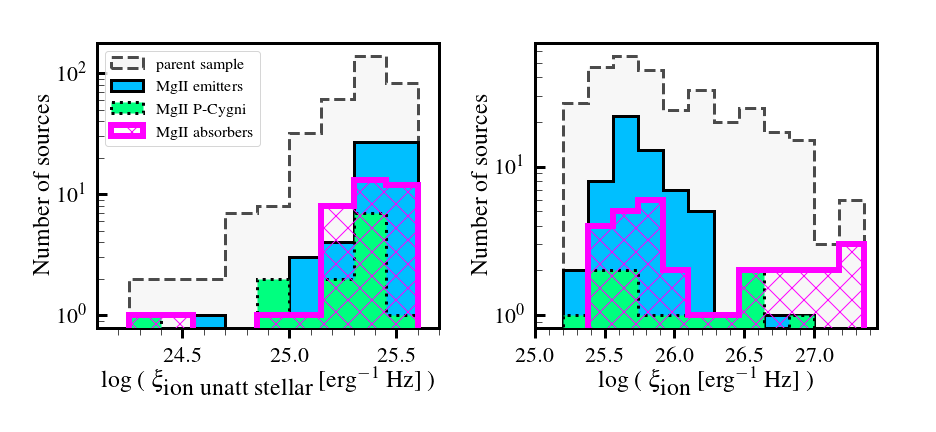}
   \caption{Ionizing emissivity computed on the unattenuated {\it stellar} (left) and dust attenuated (right) UV luminosity, $\xi_{\mbox{ion}\, stellar}$ and  $\xi_{\mbox{ion}}$, respectively. Histograms are color-coded as labelled in the legend. }
            \label{Fig9}%
   \end{figure*}

\begin{enumerate}[(i)]
\item  $\xi_{\mbox{ion}\, stellar}$ computed using only the unattenuated {\itshape stellar} UV luminosity, i.e. ignoring the absorption and re-emission of photons inside the photo-ionization regions and dust attenuation;
\item $\xi_{\mbox{ion}}$ computed using the total attenuated UV flux,  i.e. the photons that remain after the transfer of the stellar radiation through \hii\ regions and interstellar dust, across a $100 \AA$ window centred at $\lambda = 1500 \AA$.
\end{enumerate}

Fig. \ref{Fig9} shows the distributions of the two ionizing emissivities, $\xi_{\mbox{ion}\, stellar}$ and $\xi_{\mbox{ion}}$, for the whole sample. 
\mgii\ absorbers and emitters have similar distributions of the ionizing emissivity $\xi_{\mbox{ion}\, stellar}$ (left panel) computed on the unattenuated flux.
When accounting for dust attenuation, \mgii\ emitters and absorbers show different distributions of the ionizing emissivity $\xi_{\mbox{ion}}$ (right panel),
with \mgii\ emitters strongly peaking at lower values of $\xi_{\mbox{ion}}$. 
\mgii\ absorbers, on the other hand, reach higher values of $\xi_{\mbox{ion}}$.
The errors associated with the $\xi_{\mbox{ion}}$ values are of the order of $0.1-0.2$ dex, which is relatively large considered the small range of values concerned (Fig. \ref{Fig9}).
This is because the fit is, in most of the cases, constrained by the HST broad-band continuum and few emission lines.
It is worth highlighting that the $\xi_{\mbox{ion}}$ values shown here are not directly comparable with the values derived purely from nebular emission lines.
$\xi_{\mbox{ion}}$ is commonly estimated from (dust corrected) hydrogen recombination lines \citep[e.g.][]{Bouwens2016, Schaerer2016, Matthee2017, Harikane2018,Shivaei2018} and, alternatively, by exploiting UV emission lines and photoionization models \citep[e.g.]{Stark2015b,Stark2017,Nakajima2018b}. 
Additional lines are required to further constrain this physical quantity for our sample. 
Moreover, the productions of ionizing photons is connected with the properties of the stellar populations (age, metallicity and inclusion or not of binary stars) and with the escape fraction of ionizing photons 
\citep[see section 4 of][for a discussion of the uncertainties related to the interpretation of $\xi_{\mbox{ion}}$]{Nakajima2018b}.
We focus here on a qualitative comparison of the distributions of $\xi_{\mbox{ion}\, stellar}$ and $\xi_{\mbox{ion}}$ which mainly reflect the different dust attenuation experienced by \mgii\ emitters and absorbers.

The ionizing emissivity depend on the age and metallicity
of the stellar populations. We did not find differences in the age and metallicity distributions of the \mgii\ emitters and absorbers, probably because of the relatively large
redshift range explored here and the high number of free parameters into play. 
In the case of an equal release of ionizing photons, the ionizing emissivity depends only on the UV luminosity. 
Since the intrinsic ({\it stellar}) emissivity, $\xi_{\mbox{ion}\, stellar}$, is similar for \mgii\ emitters and absorbers,  
the higher $\xi_{\mbox{ion}}$ reached by the \mgii\ absorbers can be explained in terms of lower (i.e., more attenuated by dust) observed UV luminosity.
This implies that the ionizing source (i.e. stellar) in \mgii\ absorbers and emitters is intrinsically 
similar and that the differences between the two are mainly due to different dust and neutral gas content in the galaxy ISM.

\section{Discussion}\label{sec:discussion}

\subsection{Predictions from photoionization models}\label{sec:disc_models}

Early theoretical works already predicted emission from the \mgii\ doublet 
in gaseous nebulae \citep[e.g.][and references therein]{Gurzadyan1997}.
Indeed, \mgi\ is relatively easy to ionize, given its low ionization potential of  $\sim 7.65$ eV, 
and electron collisions are efficient due to the small excitation potential of the \mgii\ resonant level ($\sim 4.4$ eV). 
Dust can lead to a decrement of the emission line fluxes through absorption of photons.
The associated photon scattering, within the ionized and neutral ISM, can give rise to the absorption features observed in galaxy spectra. 
The photoionization models of \cite{Gutkin2016} account for dust attenuation and resonant scattering effects within the \hii\ regions, as 
\cloudy\ uses a full treatment of optical depths and collisional excitation for multiple lines, including \mgii.

\begin{figure}
  \centering
   \includegraphics[width=8cm]{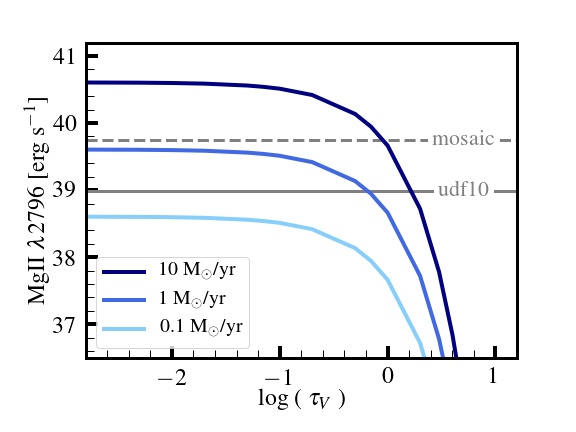}
   \caption{Predictions of the \mgiib\ line strength as function of the galaxy dust attenuation, expressed in terms of the optical depth in the V-band. 
   These fluxes are for synthetic spectra of galaxies at $z=1$ with M$_{\star}=3\times10^{9}$ M$_{\odot}$ and SFR$=0.1, 1.0, 10.0$ M$_{\odot}/$yr$^{-1}$, as labeled in the legend. Information about the other model parameters can be found in the text. The gray horizontal lines are the 3$\sigma$ MUSE emission line flux detection limits for point-like sources at $2800\,\AA$ (rest-frame) for \mosaic\ and \udften, dashed and continuous lines, respectively.}
            \label{Fig10}%
   \end{figure}

Fig. \ref{Fig10} shows the impact of pure dust attenuation (i.e. without including resonant scattering from the neutral ISM) beyond the \hii\ regions
on the \mgiib\ intensities predicted by the models (solid lines).
The predictions are for synthetic spectra of galaxies at $z=1$ with M$_{\star}=3\times10^{9}$ M$_{\odot}$ and three values of the SFR$=0.1, 1.0, 10.0 $ M$_{\odot}/$yr$^{-1}$ (color-coded as labeled in the legend),
 computed using the photoionization models of \cite{Gutkin2016} (Sec. \ref{sec:models}) and the \beagle\ code \citep[][see also Sec. \ref{sec:beagle}]{Chevallard2016}.
For consistency with the spectral fitting setup (described in Sect. \ref{sec:fitting}), we assumed a delayed star formation history and \cite{Chabrier2003} IMF
with 100 M$_{\odot}$ as upper mass cutoff. We adopted fixed values for the metallicity ($Z=0.5$ Z$_{\sun}$), the volume averaged ionization parameter (\logU$\; = -2.0$) 
and the dust-to-metal mass ratio ($\xi_{\rm d}=0.3$), in agreement with the average values found in Sect. \ref{sec:models}. 
We applied the ISM dust attenuation model of \cite{Charlot2000} to the predicted line fluxes, shown for different values of the dust optical depth in the V-band, $\tau_{V}$, in Fig. \ref{Fig10}.

The \mgiib\ line intensity starts decreasing at optical depth $\tau_{V} \sim 0.1$, with a steeper exponential decline at $\tau_{V} \gtrsim 1.$
The horizontal lines are the 3$\sigma$ MUSE emission line flux detection limits for point-like sources \citep[see Fig. 20 of][]{Bacon2017} 
at $2800\,\AA$ (rest-frame) for \mosaic\ (dashed line) and the deeper \udften\ (continuous line).
MUSE would detect the \mgii\ emission in galaxies with SFR $\geq 1$ M$_{\odot}/$yr$^{-1}$, values consistent with Fig. \ref{Fig5} and figure 3 of \cite{Finley2017}.

The fact that we do not observe the \mgii\ emission is related, along with dust absorption, to resonant scattering effects due to an higher amount of absorbing material in the
neutral ISM. An higher amount of gas could then give rise to the \mgii\ absorption features observed in relatively massive and star-forming galaxies.
Complementary information are needed to quantify gas masses of our \mgii\ emitters and absorbers. 
Unfortunately, detections of \hi\ (tracer of the atomic gas) or CO (indirect tracer of the molecular gas H$_2$) are not available for our sample. 
We also can not constrain the gas mass from emission line fluxes as this requires the detection of the most strong optical emission lines from \oiid\ to \siid\ \citep{Brinchmann2013}
which are not fully covered by our spectra.

%-----------------------------
% Figure 10

\subsection{\mgii\ emitters versus absorbers}

Fig. \ref{Fig5} shows a clear transition from \mgii\ emission to absorption, 
in terms of stellar masses and SFR. 
Moreover, the structural analysis of HST observations has shown that \mgii\ emitters tend to have smaller sizes than absorbers \citep[see also][]{Finley2017}. 
This is consistent with the observational evidence that 
galaxy sizes increase with stellar mass \cite[e.g.][]{Shen2003,VanderWel2014}.  
Furthermore, \mgii\ emitters have been found to have bluer spectral slopes than absorbers (Fig. \ref{Fig8}), i.e. a lower dust content.
No particular trend has been found with redshift, metallicity or other nebular properties, even though these could all be likely reasons
for the observed scatter in the stellar mass$-$SFR relation shown in the middle panel of Fig. \ref{Fig5}.
As a different dust content also translates in a different amount of neutral gas, the observed \mgii\ transition from emission to absorption 
can be explained by the impact of the resonant scattering to the observed flux.
Indeed, the larger the amount of neutral gas in the ISM, the more the photons are resonantly scattered within the medium and, hence, 
the higher their probability to be absorbed by dust.

\begin{figure}
  \centering
   \includegraphics[width=8cm]{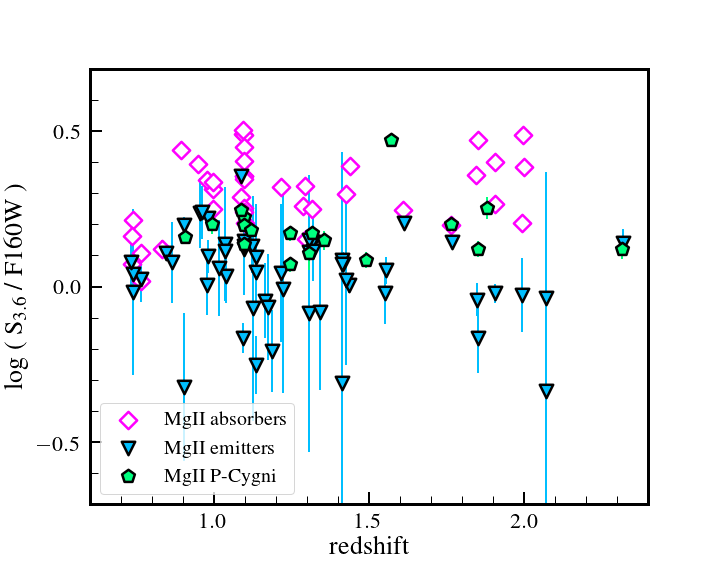}
   \caption{IRAC 3.6 $\mu$m and HST F160W flux ratio versus redshift, for the \mgii\ emitters (cyan triangle), P-Cygni (green pentagons)
   and absorbers (magenta diamonds). Error bars on \mgii\ emitters and P-Cygni are contained within the markers.}
            \label{Fig11}%
   \end{figure}

Our sources are also included in the CANDELS multi-wavelength catalogue of \citep{Guo2013}.
This catalogue combines the CANDELS \citep[Cosmic Assembly Near-infrared Deep Extragalactic Legacy Survey][]{Grogin2011,Koekemoer2011} 
HST/WFC3 F105W, F125W, and F160W data with existing public data, including the {\it Spitzer/IRAC} 3.6, 4.5, 5.8, 8.0 $\mu$m fluxes.
Fig. \ref{Fig11} shows the flux ratio from the IRAC 3.6 and HST F160W bands versus redshift.
Despite the large error bars for faint \mgii\ emitters, \mgii\ absorbers show redder F160W$-$3.6 $\mu$m colours compared to \mgii\ emitters.
The physical interpretation of the rest-frame $1-2\,\mu$m near-infrared emission from star-forming galaxies is challenging, as it depends on the nature of the stars (most likely post-AGB stars) 
contributing to it \citep[e.g.][]{Eminian2008}.
However, redder near-infrared colours have been observed in more star-forming systems, i.e. higher SFR and stellar masses \citep[e.g.][]{Rodighiero2007,Eminian2008, Mentuch2010,Lange2016}.
The redder F160W$-$3.6 $\mu$m colours observed for the \mgii\ absorbers are in agreement with their higher stellar masses and SFR, as shown in Fig. \ref{Fig5}.

\subsection{\mgii\ as an analogy to Ly$\alpha$}

In Fig. \ref{Fig7} we observed a trend between the \mgii\ EW and the stellar mass (left) and UV luminosity (right) of our \mgii\ emitters.
A deficit of bright \mgii\ emitters with strong emission \mgii\ EW ($\lesssim -7\AA$) is observed. 
This result is similar to findings from previous studies on the resonant Ly$\alpha$ line.
There are observational evidences \citep[e.g.][]{Ando2006,Stark2010,Furusawa2016,Hashimoto2017} for a lack of strong Ly$\alpha$ emission (EW $<-100\AA$) observed for
UV bright (hence, more massive) sources, referred to as the ``Ando-effect'' \citep{Ando2006}.
Multiple physical explanations can be invoked to describe the effect, including the neutral gas content, different dust attenuation, 
gas kinematics and age of the stellar populations \citep{Ando2006,Verhamme2008,Garel2012}.  
Alternatively, \cite{Nilsson2009} argued that the deficit of high Ly$\alpha$ EW luminous galaxies is related to the rarity of these sources. 
Both \mgii\ and Ly$\alpha$ are resonant lines, so we should expect to observe some comparable trends \citep[as argued also by][]{Erb2012,Martin2013,Rigby2014,Henry2018} 
and the previous explanations to apply to \mgii\ as well.
\cite{Guaita2011} found their faint high EW Ly$\alpha$ emitters (LAE) to have stellar masses lower than $10^{10}\,M_{\odot}$ and a lower dust content compared to 
the brighter galaxies in their sample. In addition, \cite{Guaita2011} and \cite{Hathi2016} found  Ly$\alpha$ emitters (EW $\geq$ 20 \AA) at $2<z<2.5$ to be less massive, have lower SFR and less dust.
Galaxies with strong Ly$\alpha$ emission have also been found to have smaller sizes than those with weaker emission or Ly$\alpha$ in absorption \citep[e.g.][]{Law2012,Shibuya2014,Kobayashi2016,Paulino-Afonso2018}.
These trends are similar to those observed between our fainter/low mass/less dusty/small size \mgii\ emitters and brighter/more massive/dust-rich/larger size \mgii\ absorbers.
In this respect, our results suggest that the lack of bright (and massive) \mgii\ emitters with strong emission \mgii\ EW (highly negative EW values) could be 
 associated with an increased amount of the ISM within the galaxy itself, invoked to explain similar trends for the Ly$\alpha$ \citep[e.g.][]{Hashimoto2017}.

%-----------------------------
% Figure 11

\subsection{\mgii\ escape fraction}\label{sec:fesc}

%-----------------------------
% Figure 12
\begin{figure}
 \centering
  \includegraphics[width=9.5cm]{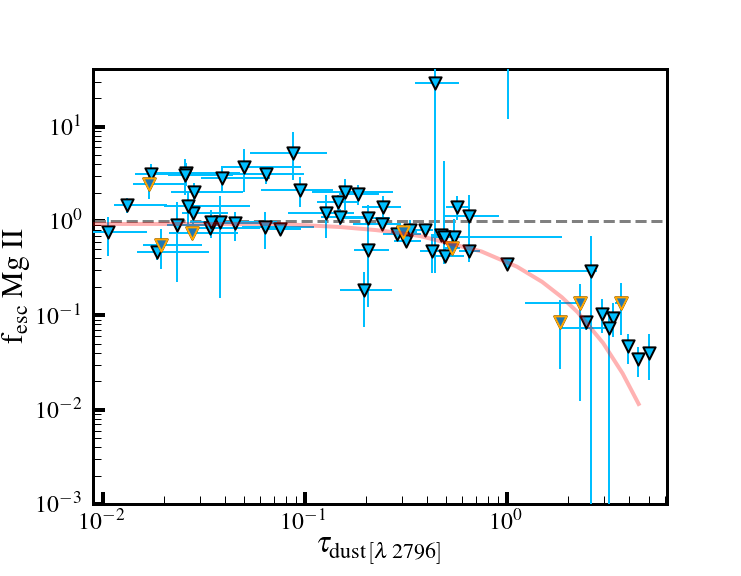}
   \caption{\mgii\ escape fraction as function of the optical depth at the \mgii $\lambda2796$ wavelength for our \mgii\ emitters. Orange contoured triangles are 
   \mgii\ emitters at $z<0.86$. The salmon curve shows the \cite{Calzetti2000} attenuation curve.}
            \label{Fig12}%
  \end{figure}

\mgii\ photons within H {\sc ii} regions can resonantly scatter through the surrounding medium until they are absorbed by dust.
The fraction of \mgii\ photons that escape the galaxy, $f_{\rm esc} \,(\mgii)$, provides important clues on the impact of dust on resonant lines.
This fraction is defined as 
the ratio between the \mgii\ observed flux and the intrinsic \mgii\ nebular emission \citep[analogous to the escape fraction of Ly$\alpha$,][]{Verhamme2008,Hayes2010, Hayes2011,Blanc2011, Atek2014}. 
This ratio is equal to unity in absence of absorption; the lower its value, the higher is
the number of photons absorbed by dust grains. 

From a study of 10 local Green Pea galaxies, \cite{Henry2018} found the escape fraction 
of \mgii\ photons to closely correlate with that of Ly$\alpha$ photons, suggesting that
both lines arise from the same low column density ISM gas.
\cite{Henry2018} proposed a prescription, calibrated on photoionization models, to predict the intrinsic \mgii\ emission
 from the \oii\ and \oiii\ line fluxes.
Unfortunately, the MUSE spectral coverage does not enable us to observe simultaneously both lines in our studied redshift range. 
Out of our 63 \mgii\ emitters, only 8 spectra exhibit both \oii\ and \oiii.

We estimated the \mgii\ escape fraction of our \mgii\ emitters from the ratio between the \mgiib fluxes, computed using \platefit\ (i.e. including correction for stellar absorption),
and the \mgiib nebular flux from the photoionization models included in the \beagle\ fit.
The observed \mgiib fluxes are dereddened using the optical depth inferred by the \beagle\ fit.
Results are illustrated in Fig. \ref{Fig12}. 
For comparison purpose, we also show the escape fraction computed assuming a \cite{Calzetti2000} attenuation curve.
The $f_{\rm esc} \,$(\mgii) is lower than unity within errors for 73\% (46/63) of our sample.
For the 38/63 sources with $f_{\rm esc} \,(\mgii) < 1$, the escape fraction ranges from 3\% to 98\%, with a median value of $\sim$$54\%$.
We note that \mgii\ emitters with dust optical depths, $\tau_{\rm d\, 2796}$,
higher than unity have the lowest values of \mgii\ escape fraction, 
resembling the observed decrease of the Ly$\alpha$ escape fraction with reddening E(B-V) \citep[e.g.][]{Hayes2010,Hayes2011,Blanc2011,Atek2014}. 
Despite the large number of variables that control the computation of $f_{\rm esc}$, as discussed below, our results
lie around the dust attenuation curve of \cite{Calzetti2000} which corresponds to pure dust attenuation. This suggests little effect by resonant scattering on the observed fluxes of our \mgii\ emitters and therefore a low content of neutral gas.
In addition, by comparing the $f_{\rm esc} \,$(\mgii) with the observed EW of \mgiib\ and the intrinsic nebular \mgiib\ flux from photoionization models, we found trends similar to those observed by \citep[][see also their Figure 5]{Henry2018}.  
Specifically, high $f_{\rm esc} \,$(\mgii) is not always associated with a strong observed \mgii\ emission. The strongest \mgii\ emitters (in terms of \mgiib\ EW) of our sample are those with the weakest intrinsic nebular \mgiib\ flux. 
These findings are in agreement with the interpretation by \cite{Henry2018} that strong \mgii\ emission is not uniquely associated with an higher production of \mgii\  photons but rather it is observed when \mgii\ photons can escape the ISM.

The reasons for observing $f_{\rm esc} \,$(\mgii) higher than unity are multifold, arising from both uncertainties on the observed quantities and assumptions on the modelling approach.

Firstly, the spectral fitting procedure described in Sec. \ref{sec:fitting} is complex as it aims at reproducing the full SED of our galaxies, accounting simultaneously for the continuum (from stars and gas) and nebular emissions. 
The HST photometric bands provide strong constraints on the choice of models, in particular in the cases where few emission lines are detected.
For instance, 7 of the sources (at $z>1.49$) for which the only line detected is the \ciii\ doublet,
have $f_{\rm esc} \,(\mgii) > 1$. In contrast, the majority of the sources (at $z<0.86$) with the highest number of emission lines detected,
including \oiiibpt, have $f_{\rm esc} \,(\mgii) < 1$ (orange contoured triangles in Fig. \ref{Fig12}).
Clearly, the computation of the \mgii\ escape fraction benefits
from additional information on optical emission lines, to 
better constrain the nebular properties (as also discussed in Sec. \ref{sec:others}). 

A second reason is related to a possible extended \mgii\ emission, similarly to that observed for Ly$\alpha$ \citep[e.g.][]{Herenz2015,Wisotzki2016, Drake2017,Leclercq2017, Marino2018}
and \ion{Fe}{ii}* \citep{Finley2017a}.
If this is case, the observed emission line ratios might be altered by the approach chosen to estimate the line flux \citep[see Sect. 4.1 of ][]{Drake2017}.
Furthermore, given the low ionization and excitation potential of \mgii\ ($\sim$7.65 eV and $\sim$4.4 eV, respectively), its emission might extend beyond the outer radius of the 
photoionization calculations\footnote{The \cloudy\ calculations in the \cite{Gutkin2016} models stop when the electron density falls below 1\% of
the hydrogen density or if the temperature falls below 100 K.}. 
Effects related to shocks, compression and turbulence within the medium could also provide an additional contribution to the \mgii\ observed flux. 
Geometrical effects, i.e. the angular distribution of \mgii\ photons due to resonant scattering, are possibly another contributing factor.
Follow up studies should, therefore, focus on modelling the emission of resonant low-ionization lines beyond the hydrogen
ionization front and on comparing the spatially resolved emission from \mgii\ to that of other non-resonant lines, e.g. \oii\ and \ciii\ \citep[see also][]{Martin2013}.

It is also worth mentioning that \mgii\ is a refractory element and, hence, sensitive to 
the level of depletion onto dust grains \citep{Guseva2013}.  
In the \cite{Gutkin2016}  models, $\sim$20\% up to $\sim80\%$ of \mgii\ is confined in dust. 
This range is consistent with $\sim$50\% of magnesium depleted onto dust grains found by \cite{Guseva2013}.
Hence, it is unlikely a major explanation for $f_{\rm esc} \,(\mgii) >1$. 
However, this, and any other dependence on model assumptions, needs to be taken 
into account when comparing observations with theoretical predictions. This is particularly true when studying 
high redshift sources, as their dust properties can differ from the local ones \citep[][]{Cucciati2012} and models might require a finer tuning.
    
Finally, we note that \platefit\ uses templates \citep[i.e.][]{Bruzual&Charlot2003} to subtract the stellar continuum different than those used in \beagle\ (i.e. Bruzual \& Charlot, in prep.). 
In their work on low redshift sources, \cite{Guseva2013} applied a constant correction for the stellar continuum to the observed fluxes (by multiplying the line intensities for $(EW + 0.5)/EW$),
motivated by the roughly constant \mgii\ stellar features in the models \citep[i.e.][]{Bruzual&Charlot2003} over a wide range of stellar ages.
Similarly, \cite{Henry2018} applied a 0.3 \AA correction to the EW of each \mgii\ doublet component.
Given the small extent of this correction, we expect the differences among templates to 
play a secondary role.

To conclude, we caution any quantitative study of the \mgii\ escape fraction, given the complexity, both from 
the observational and theoretical side, in controlling its estimate. However, even if further work is required, we emphasize the potential use of \mgii\ as complementary and, 
hopefully, alternative, resonant line to the more well studied Ly$\alpha$.
Indeed, even if much fainter, \mgii\ is less affected by the absorption of the intergalactic medium than Ly$\alpha$ \citep{Henry2018}.
   
\section{Summary and Conclusions}\label{sec:conclusions}

We have studied a sample of 381 galaxies
in the MUSE {\it Hubble} Ultra Deep Field Survey \citep{Bacon2017}
within a redshift range of $0.7<z<2.34$ .
Out of these, 123 galaxies show a wide variety of \mgiit\ doublet profiles,
ranging from emission to absorption,
and have been classified in: 
\mgii\ emitters (63), \mgii\ P-Cygni (19) and \mgii\ absorbers (41).
The main goal of this work was to find the underlying physical properties that drive the detection of the \mgii\ doublet in emission or in absorption 
in the spectra of galaxies up to the peak of the SFRD history \citep{Madau2014}, where a change in the regime of the
global properties \citep[also in term of dust content,][]{Cucciati2012} of star-forming galaxies is observed. 
In what follows, we summarize the main results of this work and highlight future prospectives.

\begin{itemize}
\item[$\bullet$] We first explored whether the line fluxes of \mgii\ emitters
 are consistent with predictions from nebular emission in star-forming galaxies (Fig. \ref{Fig3}) and find them
 to be compatible with ionizing photons produced within \hii\ regions \citep[see also][]{Erb2012}.

\item[$\bullet$] Based on the measured emission line ratios available in our sample (i.e. \oii/\mgii, \neiii/\mgii\ and \ciii/\mgii),
we were not able to identify features that would enable us to confirm or reject AGN or radiative shocks as sources of ionization.
However, we removed all the potential AGN identified through X-ray from our sample (see Sect. \ref{sec:sample} and \ref{sec:others})
 and we do not expect our \mgii\ emitters to be contaminated by strong AGN. 
While we can dismiss a substantial contribution from gravitational accretion onto black holes,
we can not rule out a possible contamination from shocks to the observed spectra.

\item[$\bullet$] To explore whether the different profiles of the \mgii\ doublet trace particular properties 
of the galaxies, we relied on the Bayesian spectral fitting code \beagle\ \citep{Chevallard2016}. 
Specifically, we exploited both the information from MUSE spectra and HST broad-band photometry 
and performed a spectral fit accounting simultaneously for the continuum and nebular stellar emissions (examples are shown in Fig. \ref{Fig4}).
We found \mgii\ emitters to have, on average, lower stellar masses and bluer spectral slopes (i.e. lower dust attenuation)
than \mgii\ absorbers (Figs. \ref{Fig5}, \ref{Fig8} and \ref{Fig9}). 
This indicates that the content of dust and neutral gas in the ISM of our galaxies plays a dominant role 
in shaping the \mgii\ features. 

\item[$\bullet$] We found a lack of strong \mgii\ EW in emission for our bright \mgii\ emitters (Fig. \ref{Fig7}), similarly to the trend that has been observed for Ly$\alpha$.
As they are both resonant lines, one would expect to find correlations between the observed properties of \mgii\ and Ly$\alpha$, upon the assumptions
that the source of photons production is the same and that they are observed through the same medium. 

\item[$\bullet$] We compute the fraction of \mgii\ photons escaping the galaxy (Fig. \ref{Fig12}) finding many uncertainties
in deriving this quantity. These uncertainties are mainly related to our 
modelling approach and other physical processes that could come into play (e.g. shocks, geometrical effects).
Additional spectral information (preferentially spatially resolved) and a tailored modelling of the \mgii\ resonant doublet are necessary to 
fully understand its emission feature. 
Nevertheless, extending the escape fraction studies, until now mainly restricted to the Ly$\alpha$ 
\citep[e.g.][]{Verhamme2008,Hayes2010, Hayes2011,Blanc2011, Atek2014}, 
with other resonant lines, such as \mgii, is a promising approach to better understand the physics of the ISM within galaxies \citep[see also][]{Henry2018}.
\end{itemize}

Additional emission lines in the rest-frame UV/optical range, along with 
deep spatially resolved spectroscopy, are required to further investigate the
physical origin of \mgii\ emission. 
This is specially relevant for those sources with
an observed \mgii\ flux higher than that predicted by purely nebular emission and 
for studying the potential effect of radiative shocks 
on the observed spectra of our \mgii\ emitters. 
In particular, the completion of deeper MUSE 
observations with adaptive optics, planned for a subregion of 1 square arcminute of the UDF, will enable further studies in this direction.
We expect new data for at least 34/381 galaxies of our sample (including 6/2/2 \mgii\ emitters/P-Cygni/absorbers).

Most likely, a contribution from shocks is present in those sources with \mgii\ P-Cygni-like profiles, 
as this peculiar profile is a tracer of galactic winds. A detailed study of the \mgii\ spectral shape
and physical properties of these sources is beyond the scope of this paper and will be subject of future analysis.
It is worth highlighting that galaxies with \mgii\ P-Cygni profile show intermediate properties between emitters and absorbers. 
Additional studies on \mgii\ P-Cygni will help us understand whether
\mgii\ emitters are the results of a complete ``blowout'' of gas from galactic winds 
or if their stellar populations have simply fully ionized the gas. 
In this respect, we note that 6 of our \mgii\ emitters (ID 84, 121, 46, 92, 1219, 7101) are in major close pairs \citep[Table 3 of][]{Ventou2017}, 
pointing to major mergers. These merging events may have indeed produced galactic outflows for these sources \citep[see also][]{Finley2017}. 
Follow up studies will focus on explaining the observed transitions between \mgii\ in emission, through P-Cygni profile, 
to absorption by means of proper radiative transfer modelling of the neutral medium (Garel et al., in prep., Finley et al., in prep.).

Our spectral modelling accounts for photospheric stellar emission and absorption and for the nebular emission from
ionized gas, but does not include the effect of the neutral ISM. Spectral models incorporating absorption by neutral gas
have been developed by \cite{Vidal2017} and are now being implemented wihin the fitting code \beagle\ to get a 
more comprehensive and simultaneous treatment of the spectral features from stars, ionized and neutral ISM in and around galaxies.

To conclude, pushing these studies to higher redshift will allow us to see whether \mgii\ emitters are common or rare at earlier epochs.
The \mgii\ line doublet is also very promising because it undergoes the same physics as the Ly$\alpha$ line, due to their resonant nature.
A close connection between the profiles of these two lines has been already found in Green Peas galaxies at $z\sim 0.2-0.3$ \citep{Henry2018}.
It would be extremely interesting to see whether a connection exist among the properties of these two lines up to earlier epochs of galaxy evolution.
A possibility is to search for \mgii\ emission in near-infrared spectroscopic observations complementary those of the MUSE LAE. 
For example, K-band Multi-Object Spectrometer (KMOS) observations are available for the MUSE-Wide LAE  and 3D-HST grism spectroscopy for the MUSE UDF.
Moreover, future near-infrared spectrographs, such as JWST/NIRSpec, will also potentially enable the studies of the \mgii\ features at higher redshifts than those 
explored in this work, along with the simultaneous observations of Ly$\alpha$ for the most distant sources. 
This will provide additional clues on the dust and gas content of the ISM at high redshifts and, why not, during the reionization epoch.

\begin{acknowledgements} 
AF thanks Emma Curtis-Lake and Alba Vidal-Garc\'ia for valuable input on SED modelling and fitting.
This work is supported  by the ERC  advanced  grant  339659-MUSICOS  (R.  Bacon). 
This work makes use of the {\it MUSE Python Data Analysis Framework}, MPDAF ({\url{https://git-cral.univ-lyon1.fr/MUSE/mpdaf}}), developed at the Centre de Recherche Astrophysique de Lyon (CRAL).
LT and AF thanks Baptiste Pellorce for his contribution during his UCBL1 undergrade internship at CRAL.
DC and JR acknowledge support from the ERC starting grant  336736-CALENDS.
SC and JC acknowledge financial support from the European Research Council via an Advanced Grant under grant agreement no. 321323 (NEOGAL).
TC acknowledges support of the ANR FOGHAR (ANR-13-BS05-0010-02), the OCEVU Labex (ANR-11-LABX-0060), and the A*MIDEX project (ANR-11-IDEX-0001-02) funded by the "Investissements d'avenir" French government program. 
JB acknowledges support by Funda{\c c}{\~a}o para a Ci{\^e}ncia e a Tecnologia (FCT) through national funds (UID/FIS/04434/2013) and by FEDER through COMPETE2020 (POCI-01-0145-FEDER-007672) and Investigador FCT contract IF/01654/2014/CP1215/CT0003.

\end{acknowledgements}

% for the bibliography, at the end
\bibliographystyle{aa} % style aa.bst
\bibliography{biblio_MgII} % your references Yourfile.bib

\begin{thebibliography}{132}
\expandafter\ifx\csname natexlab\endcsname\relax\def\natexlab#1{#1}\fi

\bibitem[{{Aguirre} {et~al.}(2001){Aguirre}, {Hernquist}, {Schaye}, {Katz},
  {Weinberg}, \& {Gardner}}]{Aguirre2001}
{Aguirre}, A., {Hernquist}, L., {Schaye}, J., {et~al.} 2001, \apj, 561, 521

\bibitem[{{Ali} {et~al.}(1991){Ali}, {Blum}, {Bumgardner}, {Cranmer},
  {Ferland}, {Haefner}, \& {Tiede}}]{Ali1991}
{Ali}, B., {Blum}, R.~D., {Bumgardner}, T.~E., {et~al.} 1991, \pasp, 103, 1182

\bibitem[{{Allen} {et~al.}(2008){Allen}, {Groves}, {Dopita}, {Sutherland}, \&
  {Kewley}}]{Allen2008}
{Allen}, M.~G., {Groves}, B.~A., {Dopita}, M.~A., {Sutherland}, R.~S., \&
  {Kewley}, L.~J. 2008, \apjs, 178, 20

\bibitem[{{Ando} {et~al.}(2006){Ando}, {Ohta}, {Iwata}, {Akiyama}, {Aoki}, \&
  {Tamura}}]{Ando2006}
{Ando}, M., {Ohta}, K., {Iwata}, I., {et~al.} 2006, \apjl, 645, L9

\bibitem[{{Atek} {et~al.}(2014){Atek}, {Kneib}, {Pacifici}, {Malkan},
  {Charlot}, {Lee}, {Bedregal}, {Bunker}, {Colbert}, {Dressler}, {Hathi},
  {Lehnert}, {Martin}, {McCarthy}, {Rafelski}, {Ross}, {Siana}, \&
  {Teplitz}}]{Atek2014}
{Atek}, H., {Kneib}, J.-P., {Pacifici}, C., {et~al.} 2014, \apj, 789, 96

\bibitem[{{Bacon} {et~al.}(2015){Bacon}, {Brinchmann}, {Richard}, {Contini},
  {Drake}, {Franx}, {Tacchella}, {Vernet}, {Wisotzki}, {Blaizot}, {Bouch{\'e}},
  {Bouwens}, {Cantalupo}, {Carollo}, {Carton}, {Caruana}, {Cl{\'e}ment},
  {Dreizler}, {Epinat}, {Guiderdoni}, {Herenz}, {Husser}, {Kamann}, {Kerutt},
  {Kollatschny}, {Krajnovic}, {Lilly}, {Martinsson}, {Michel-Dansac},
  {Patricio}, {Schaye}, {Shirazi}, {Soto}, {Soucail}, {Steinmetz}, {Urrutia},
  {Weilbacher}, \& {de Zeeuw}}]{Bacon2015}
{Bacon}, R., {Brinchmann}, J., {Richard}, J., {et~al.} 2015, \aap, 575, A75

\bibitem[{{Bacon} {et~al.}(2017){Bacon}, {Conseil}, {Mary}, {Brinchmann},
  {Shepherd}, {Akhlaghi}, {Weilbacher}, {Piqueras}, {Wisotzki}, {Lagattuta},
  {Epinat}, {Guerou}, {Inami}, {Cantalupo}, {Courbot}, {Contini}, {Richard},
  {Maseda}, {Bouwens}, {Bouch{\'e}}, {Kollatschny}, {Schaye}, {Marino},
  {Pello}, {Herenz}, {Guiderdoni}, \& {Carollo}}]{Bacon2017}
{Bacon}, R., {Conseil}, S., {Mary}, D., {et~al.} 2017, \aap, 608, A1

\bibitem[{{Berg} {et~al.}(2018){Berg}, {Erb}, {Auger}, {Pettini}, \&
  {Brammer}}]{Berg2018}
{Berg}, D.~A., {Erb}, D.~K., {Auger}, M.~W., {Pettini}, M., \& {Brammer}, G.~B.
  2018, ArXiv e-prints [\eprint[arXiv]{1803.02340}]

\bibitem[{{Berg} {et~al.}(2016){Berg}, {Skillman}, {Henry}, {Erb}, \&
  {Carigi}}]{Berg2016}
{Berg}, D.~A., {Skillman}, E.~D., {Henry}, R.~B.~C., {Erb}, D.~K., \& {Carigi},
  L. 2016, \apj, 827, 126

\bibitem[{{Blanc} {et~al.}(2011){Blanc}, {Adams}, {Gebhardt}, {Hill}, {Drory},
  {Hao}, {Bender}, {Ciardullo}, {Finkelstein}, {Fry}, {Gawiser}, {Gronwall},
  {Hopp}, {Jeong}, {Kelzenberg}, {Komatsu}, {MacQueen}, {Murphy}, {Roth},
  {Schneider}, \& {Tufts}}]{Blanc2011}
{Blanc}, G.~A., {Adams}, J.~J., {Gebhardt}, K., {et~al.} 2011, \apj, 736, 31

\bibitem[{{Boogard} {et~al.}(2018){Boogard}, {Brinchmann}, {Bouch{\'e}}, M., \&
  R.}]{Boogaard2018}
{Boogard}, L., {Brinchmann}, J., {Bouch{\'e}}, N., M., P., \& R., B. 2018,
  submitted to \aap

\bibitem[{{Bordoloi} {et~al.}(2016){Bordoloi}, {Rigby}, {Tumlinson}, {Bayliss},
  {Sharon}, {Gladders}, \& {Wuyts}}]{Bordoloi2016}
{Bordoloi}, R., {Rigby}, J.~R., {Tumlinson}, J., {et~al.} 2016, \mnras, 458,
  1891

\bibitem[{{Bouwens} {et~al.}(2016){Bouwens}, {Smit}, {Labb{\'e}}, {Franx},
  {Caruana}, {Oesch}, {Stefanon}, \& {Rasappu}}]{Bouwens2016}
{Bouwens}, R.~J., {Smit}, R., {Labb{\'e}}, I., {et~al.} 2016, \apj, 831, 176

\bibitem[{{Brammer} {et~al.}(2012){Brammer}, {van Dokkum}, {Franx},
  {Fumagalli}, {Patel}, {Rix}, {Skelton}, {Kriek}, {Nelson}, {Schmidt},
  {Bezanson}, {da Cunha}, {Erb}, {Fan}, {F{\"o}rster Schreiber}, {Illingworth},
  {Labb{\'e}}, {Leja}, {Lundgren}, {Magee}, {Marchesini}, {McCarthy},
  {Momcheva}, {Muzzin}, {Quadri}, {Steidel}, {Tal}, {Wake}, {Whitaker}, \&
  {Williams}}]{Brammer2012}
{Brammer}, G.~B., {van Dokkum}, P.~G., {Franx}, M., {et~al.} 2012, \apjs, 200,
  13

\bibitem[{{Bressan} {et~al.}(2012){Bressan}, {Marigo}, {Girardi}, {Salasnich},
  {Dal Cero}, {Rubele}, \& {Nanni}}]{Bressan2012}
{Bressan}, A., {Marigo}, P., {Girardi}, L., {et~al.} 2012, \mnras, 427, 127

\bibitem[{{Brinchmann} {et~al.}(2013){Brinchmann}, {Charlot}, {Kauffmann},
  {Heckman}, {White}, \& {Tremonti}}]{Brinchmann2013}
{Brinchmann}, J., {Charlot}, S., {Kauffmann}, G., {et~al.} 2013, \mnras, 432,
  2112

\bibitem[{{Brinchmann} {et~al.}(2004){Brinchmann}, {Charlot}, {White},
  {Tremonti}, {Kauffmann}, {Heckman}, \& {Brinkmann}}]{Brinchmann2004}
{Brinchmann}, J., {Charlot}, S., {White}, S.~D.~M., {et~al.} 2004, \mnras, 351,
  1151

\bibitem[{{Brinchmann} {et~al.}(2008){Brinchmann}, {Pettini}, \&
  {Charlot}}]{Brinchmann2008}
{Brinchmann}, J., {Pettini}, M., \& {Charlot}, S. 2008, \mnras, 385, 769

\bibitem[{{Bruzual} \& {Charlot}(2003)}]{Bruzual&Charlot2003}
{Bruzual}, G. \& {Charlot}, S. 2003, \mnras, 344, 1000

\bibitem[{{Calzetti} {et~al.}(2000){Calzetti}, {Armus}, {Bohlin}, {Kinney},
  {Koornneef}, \& {Storchi-Bergmann}}]{Calzetti2000}
{Calzetti}, D., {Armus}, L., {Bohlin}, R.~C., {et~al.} 2000, \apj, 533, 682

\bibitem[{{Calzetti} {et~al.}(1994){Calzetti}, {Kinney}, \&
  {Storchi-Bergmann}}]{Calzetti1994}
{Calzetti}, D., {Kinney}, A.~L., \& {Storchi-Bergmann}, T. 1994, \apj, 429, 582

\bibitem[{{Chabrier}(2003)}]{Chabrier2003}
{Chabrier}, G. 2003, \pasp, 115, 763

\bibitem[{{Charlot} \& {Fall}(2000)}]{Charlot2000}
{Charlot}, S. \& {Fall}, S.~M. 2000, \apj, 539, 718

\bibitem[{{Charlot} \& {Longhetti}(2001)}]{Charlot&Longhetti2001}
{Charlot}, S. \& {Longhetti}, M. 2001, \mnras, 323, 887

\bibitem[{{Chevallard} \& {Charlot}(2016)}]{Chevallard2016}
{Chevallard}, J. \& {Charlot}, S. 2016, \mnras, 462, 1415

\bibitem[{{Chevallard} {et~al.}(2017){Chevallard}, {Charlot}, {Senchyna},
  {Stark}, {Vidal-Garc{\'{\i}}a}, {Feltre}, {Gutkin}, {Jones}, {Mainali}, \&
  {Wofford}}]{Chevallard2017}
{Chevallard}, J., {Charlot}, S., {Senchyna}, P., {et~al.} 2017, ArXiv e-prints
  [\eprint[arXiv]{1709.03503}]

\bibitem[{{Chevallard} {et~al.}(2013){Chevallard}, {Charlot}, {Wandelt}, \&
  {Wild}}]{Chevallard2013}
{Chevallard}, J., {Charlot}, S., {Wandelt}, B., \& {Wild}, V. 2013, \mnras,
  432, 2061

\bibitem[{{Cucciati} {et~al.}(2012){Cucciati}, {Tresse}, {Ilbert}, {Le
  F{\`e}vre}, {Garilli}, {Le Brun}, {Cassata}, {Franzetti}, {Maccagni},
  {Scodeggio}, {Zucca}, {Zamorani}, {Bardelli}, {Bolzonella}, {Bielby},
  {McCracken}, {Zanichelli}, \& {Vergani}}]{Cucciati2012}
{Cucciati}, O., {Tresse}, L., {Ilbert}, O., {et~al.} 2012, \aap, 539, A31

\bibitem[{{Drake} {et~al.}(2017){Drake}, {Guiderdoni}, {Blaizot}, {Wisotzki},
  {Herenz}, {Garel}, {Richard}, {Bacon}, {Bina}, {Cantalupo}, {Contini}, {den
  Brok}, {Hashimoto}, {Marino}, {Pell{\'o}}, {Schaye}, \&
  {Schmidt}}]{Drake2017}
{Drake}, A.~B., {Guiderdoni}, B., {Blaizot}, J., {et~al.} 2017, \mnras, 471,
  267

\bibitem[{{Eminian} {et~al.}(2008){Eminian}, {Kauffmann}, {Charlot}, {Wild},
  {Bruzual}, {Rettura}, \& {Loveday}}]{Eminian2008}
{Eminian}, C., {Kauffmann}, G., {Charlot}, S., {et~al.} 2008, \mnras, 384, 930

\bibitem[{{Erb} {et~al.}(2012){Erb}, {Quider}, {Henry}, \& {Martin}}]{Erb2012}
{Erb}, D.~K., {Quider}, A.~M., {Henry}, A.~L., \& {Martin}, C.~L. 2012, \apj,
  759, 26

\bibitem[{{Feltre} {et~al.}(2016){Feltre}, {Charlot}, \& {Gutkin}}]{Feltre2016}
{Feltre}, A., {Charlot}, S., \& {Gutkin}, J. 2016, \mnras, 456, 3354

\bibitem[{{Ferland} {et~al.}(2013){Ferland}, {Porter}, {van Hoof}, {Williams},
  {Abel}, {Lykins}, {Shaw}, {Henney}, \& {Stancil}}]{Ferland2013}
{Ferland}, G.~J., {Porter}, R.~L., {van Hoof}, P.~A.~M., {et~al.} 2013, \rmxaa,
  49, 137

\bibitem[{{Finlator} \& {Dav{\'e}}(2008)}]{Finlator2008}
{Finlator}, K. \& {Dav{\'e}}, R. 2008, \mnras, 385, 2181

\bibitem[{{Finley} {et~al.}(2017{\natexlab{a}}){Finley}, {Bouch{\'e}},
  {Contini}, {Epinat}, {Bacon}, {Brinchmann}, {Cantalupo}, {Erroz-Ferrer},
  {Marino}, {Maseda}, {Richard}, {Schroetter}, {Verhamme}, {Weilbacher},
  {Wendt}, \& {Wisotzki}}]{Finley2017a}
{Finley}, H., {Bouch{\'e}}, N., {Contini}, T., {et~al.} 2017{\natexlab{a}},
  \aap, 605, A118

\bibitem[{{Finley} {et~al.}(2017{\natexlab{b}}){Finley}, {Bouch{\'e}},
  {Contini}, {Paalvast}, {Boogaard}, {Maseda}, {Bacon}, {Blaizot},
  {Brinchmann}, {Epinat}, {Feltre}, {Marino}, {Muzahid}, {Richard}, {Schaye},
  {Verhamme}, {Weilbacher}, \& {Wisotzki}}]{Finley2017}
{Finley}, H., {Bouch{\'e}}, N., {Contini}, T., {et~al.} 2017{\natexlab{b}},
  \aap, 608, A7

\bibitem[{{Furusawa} {et~al.}(2016){Furusawa}, {Kashikawa}, {Kobayashi},
  {Dunlop}, {Shimasaku}, {Takata}, {Sekiguchi}, {Naito}, {Furusawa}, {Ouchi},
  {Nakata}, {Yasuda}, {Okura}, {Taniguchi}, {Yamada}, {Kajisawa}, {Fynbo}, \&
  {Le F{\`e}vre}}]{Furusawa2016}
{Furusawa}, H., {Kashikawa}, N., {Kobayashi}, M.~A.~R., {et~al.} 2016, \apj,
  822, 46

\bibitem[{{Garel} {et~al.}(2012){Garel}, {Blaizot}, {Guiderdoni}, {Schaerer},
  {Verhamme}, \& {Hayes}}]{Garel2012}
{Garel}, T., {Blaizot}, J., {Guiderdoni}, B., {et~al.} 2012, \mnras, 422, 310

\bibitem[{{Giavalisco} {et~al.}(2011){Giavalisco}, {Vanzella}, {Salimbeni},
  {Tripp}, {Dickinson}, {Cassata}, {Renzini}, {Guo}, {Ferguson}, {Nonino},
  {Cimatti}, {Kurk}, {Mignoli}, \& {Tang}}]{Giavalisco2011}
{Giavalisco}, M., {Vanzella}, E., {Salimbeni}, S., {et~al.} 2011, \apj, 743, 95

\bibitem[{{Grogin} {et~al.}(2011){Grogin}, {Kocevski}, {Faber}, {Ferguson},
  {Koekemoer}, {Riess}, {Acquaviva}, {Alexander}, {Almaini}, {Ashby}, {Barden},
  {Bell}, {Bournaud}, {Brown}, {Caputi}, {Casertano}, {Cassata}, {Castellano},
  {Challis}, {Chary}, {Cheung}, {Cirasuolo}, {Conselice}, {Roshan Cooray},
  {Croton}, {Daddi}, {Dahlen}, {Dav{\'e}}, {de Mello}, {Dekel}, {Dickinson},
  {Dolch}, {Donley}, {Dunlop}, {Dutton}, {Elbaz}, {Fazio}, {Filippenko},
  {Finkelstein}, {Fontana}, {Gardner}, {Garnavich}, {Gawiser}, {Giavalisco},
  {Grazian}, {Guo}, {Hathi}, {H{\"a}ussler}, {Hopkins}, {Huang}, {Huang},
  {Jha}, {Kartaltepe}, {Kirshner}, {Koo}, {Lai}, {Lee}, {Li}, {Lotz}, {Lucas},
  {Madau}, {McCarthy}, {McGrath}, {McIntosh}, {McLure}, {Mobasher},
  {Moustakas}, {Mozena}, {Nandra}, {Newman}, {Niemi}, {Noeske}, {Papovich},
  {Pentericci}, {Pope}, {Primack}, {Rajan}, {Ravindranath}, {Reddy}, {Renzini},
  {Rix}, {Robaina}, {Rodney}, {Rosario}, {Rosati}, {Salimbeni}, {Scarlata},
  {Siana}, {Simard}, {Smidt}, {Somerville}, {Spinrad}, {Straughn}, {Strolger},
  {Telford}, {Teplitz}, {Trump}, {van der Wel}, {Villforth}, {Wechsler},
  {Weiner}, {Wiklind}, {Wild}, {Wilson}, {Wuyts}, {Yan}, \& {Yun}}]{Grogin2011}
{Grogin}, N.~A., {Kocevski}, D.~D., {Faber}, S.~M., {et~al.} 2011, \apjs, 197,
  35

\bibitem[{{Guaita} {et~al.}(2011){Guaita}, {Acquaviva}, {Padilla}, {Gawiser},
  {Bond}, {Ciardullo}, {Treister}, {Kurczynski}, {Gronwall}, {Lira}, \&
  {Schawinski}}]{Guaita2011}
{Guaita}, L., {Acquaviva}, V., {Padilla}, N., {et~al.} 2011, \apj, 733, 114

\bibitem[{{Guo} {et~al.}(2013){Guo}, {Ferguson}, {Giavalisco}, {Barro},
  {Willner}, {Ashby}, {Dahlen}, {Donley}, {Faber}, {Fontana}, {Galametz},
  {Grazian}, {Huang}, {Kocevski}, {Koekemoer}, {Koo}, {McGrath}, {Peth},
  {Salvato}, {Wuyts}, {Castellano}, {Cooray}, {Dickinson}, {Dunlop}, {Fazio},
  {Gardner}, {Gawiser}, {Grogin}, {Hathi}, {Hsu}, {Lee}, {Lucas}, {Mobasher},
  {Nandra}, {Newman}, \& {van der Wel}}]{Guo2013}
{Guo}, Y., {Ferguson}, H.~C., {Giavalisco}, M., {et~al.} 2013, \apjs, 207, 24

\bibitem[{{Gurzadyan}(1997)}]{Gurzadyan1997}
{Gurzadyan}, G.~A. 1997, {The Physics and Dynamics of Planetary Nebulae}, 179

\bibitem[{{Guseva} {et~al.}(2013){Guseva}, {Izotov}, {Fricke}, \&
  {Henkel}}]{Guseva2013}
{Guseva}, N.~G., {Izotov}, Y.~I., {Fricke}, K.~J., \& {Henkel}, C. 2013, \aap,
  555, A90

\bibitem[{{Gutkin} {et~al.}(2016){Gutkin}, {Charlot}, \&
  {Bruzual}}]{Gutkin2016}
{Gutkin}, J., {Charlot}, S., \& {Bruzual}, G. 2016, \mnras, 462, 1757

\bibitem[{{Harikane} {et~al.}(2018){Harikane}, {Ouchi}, {Shibuya}, {Kojima},
  {Zhang}, {Itoh}, {Ono}, {Higuchi}, {Inoue}, {Chevallard}, {Capak}, {Nagao},
  {Onodera}, {Faisst}, {Martin}, {Rauch}, {Bruzual}, {Charlot}, {Davidzon},
  {Fujimoto}, {Hilmi}, {Ilbert}, {Lee}, {Matsuoka}, {Silverman}, \&
  {Toft}}]{Harikane2018}
{Harikane}, Y., {Ouchi}, M., {Shibuya}, T., {et~al.} 2018, \apj, 859, 84

\bibitem[{{Harikane} {et~al.}(2014){Harikane}, {Ouchi}, {Yuma}, {Rauch},
  {Nakajima}, \& {Ono}}]{Harikane2014}
{Harikane}, Y., {Ouchi}, M., {Yuma}, S., {et~al.} 2014, \apj, 794, 129

\bibitem[{{Hashimoto} {et~al.}(2017){Hashimoto}, {Garel}, {Guiderdoni},
  {Drake}, {Bacon}, {Blaizot}, {Richard}, {Leclercq}, {Inami}, {Verhamme},
  {Bouwens}, {Brinchmann}, {Cantalupo}, {Carollo}, {Caruana}, {Herenz},
  {Kerutt}, {Marino}, {Mitchell}, \& {Schaye}}]{Hashimoto2017}
{Hashimoto}, T., {Garel}, T., {Guiderdoni}, B., {et~al.} 2017, \aap, 608, A10

\bibitem[{{Hathi} {et~al.}(2016){Hathi}, {Le F{\`e}vre}, {Ilbert}, {Cassata},
  {Tasca}, {Lemaux}, {Garilli}, {Le Brun}, {Maccagni}, {Pentericci}, {Thomas},
  {Vanzella}, {Zamorani}, {Zucca}, {Amor{\'{\i}}n}, {Bardelli}, {Cassar{\`a}},
  {Castellano}, {Cimatti}, {Cucciati}, {Durkalec}, {Fontana}, {Giavalisco},
  {Grazian}, {Guaita}, {Koekemoer}, {Paltani}, {Pforr}, {Ribeiro}, {Schaerer},
  {Scodeggio}, {Sommariva}, {Talia}, {Tresse}, {Vergani}, {Capak}, {Charlot},
  {Contini}, {Cuby}, {de la Torre}, {Dunlop}, {Fotopoulou},
  {L{\'o}pez-Sanjuan}, {Mellier}, {Salvato}, {Scoville}, {Taniguchi}, \&
  {Wang}}]{Hathi2016}
{Hathi}, N.~P., {Le F{\`e}vre}, O., {Ilbert}, O., {et~al.} 2016, \aap, 588, A26

\bibitem[{{Hayes} {et~al.}(2010){Hayes}, {{\"O}stlin}, {Schaerer}, {Mas-Hesse},
  {Leitherer}, {Atek}, {Kunth}, {Verhamme}, {de Barros}, \&
  {Melinder}}]{Hayes2010}
{Hayes}, M., {{\"O}stlin}, G., {Schaerer}, D., {et~al.} 2010, \nat, 464, 562

\bibitem[{{Hayes} {et~al.}(2011){Hayes}, {Schaerer}, {{\"O}stlin}, {Mas-Hesse},
  {Atek}, \& {Kunth}}]{Hayes2011}
{Hayes}, M., {Schaerer}, D., {{\"O}stlin}, G., {et~al.} 2011, \apj, 730, 8

\bibitem[{{Henry} {et~al.}(2018){Henry}, {Berg}, {Scarlata}, {Verhamme}, \&
  {Erb}}]{Henry2018}
{Henry}, A., {Berg}, D.~A., {Scarlata}, C., {Verhamme}, A., \& {Erb}, D. 2018,
  \apj, 855, 96

\bibitem[{{Herenz} {et~al.}(2015){Herenz}, {Wisotzki}, {Roth}, \&
  {Anders}}]{Herenz2015}
{Herenz}, E.~C., {Wisotzki}, L., {Roth}, M., \& {Anders}, F. 2015, \aap, 576,
  A115

\bibitem[{{Hirschmann} {et~al.}(2017){Hirschmann}, {Charlot}, {Feltre}, {Naab},
  {Choi}, {Ostriker}, \& {Somerville}}]{Hirschmann2017}
{Hirschmann}, M., {Charlot}, S., {Feltre}, A., {et~al.} 2017, \mnras, 472, 2468

\bibitem[{{Hummer} \& {Storey}(1987)}]{Hummer1987}
{Hummer}, D.~G. \& {Storey}, P.~J. 1987, \mnras, 224, 801

\bibitem[{{Inami} {et~al.}(2017){Inami}, {Bacon}, {Brinchmann}, {Richard},
  {Contini}, {Conseil}, {Hamer}, {Akhlaghi}, {Bouch{\'e}}, {Cl{\'e}ment},
  {Desprez}, {Drake}, {Hashimoto}, {Leclercq}, {Maseda}, {Michel-Dansac},
  {Paalvast}, {Tresse}, {Ventou}, {Kollatschny}, {Boogaard}, {Finley},
  {Marino}, {Schaye}, \& {Wisotzki}}]{Inami2017}
{Inami}, H., {Bacon}, R., {Brinchmann}, J., {et~al.} 2017, \aap, 608, A2

\bibitem[{{Izotov} {et~al.}(2016){Izotov}, {Guseva}, {Fricke}, \&
  {Henkel}}]{Izotov2016}
{Izotov}, Y.~I., {Guseva}, N.~G., {Fricke}, K.~J., \& {Henkel}, C. 2016,
  \mnras, 462, 4427

\bibitem[{{Izotov} {et~al.}(2017){Izotov}, {Guseva}, {Fricke}, {Henkel}, \&
  {Schaerer}}]{Izotov2017}
{Izotov}, Y.~I., {Guseva}, N.~G., {Fricke}, K.~J., {Henkel}, C., \& {Schaerer},
  D. 2017, \mnras, 467, 4118

\bibitem[{{Karman} {et~al.}(2016){Karman}, {Grillo}, {Balestra}, {Rosati},
  {Caputi}, {Di Teodoro}, {Fraternali}, {Gavazzi}, {Mercurio}, {Prochaska},
  {Rodney}, \& {Treu}}]{Karman2016}
{Karman}, W., {Grillo}, C., {Balestra}, I., {et~al.} 2016, \aap, 585, A27

\bibitem[{{Kewley} {et~al.}(2013){Kewley}, {Dopita}, {Leitherer}, {Dav{\'e}},
  {Yuan}, {Allen}, {Groves}, \& {Sutherland}}]{Kewley2013}
{Kewley}, L.~J., {Dopita}, M.~A., {Leitherer}, C., {et~al.} 2013, \apj, 774,
  100

\bibitem[{{Kinney} {et~al.}(1993){Kinney}, {Bohlin}, {Calzetti}, {Panagia}, \&
  {Wyse}}]{Kinney1993}
{Kinney}, A.~L., {Bohlin}, R.~C., {Calzetti}, D., {Panagia}, N., \& {Wyse},
  R.~F.~G. 1993, \apjs, 86, 5

\bibitem[{{Kobayashi} {et~al.}(2016){Kobayashi}, {Murata}, {Koekemoer},
  {Murayama}, {Taniguchi}, {Kajisawa}, {Shioya}, {Scoville}, {Nagao}, \&
  {Capak}}]{Kobayashi2016}
{Kobayashi}, M.~A.~R., {Murata}, K.~L., {Koekemoer}, A.~M., {et~al.} 2016,
  \apj, 819, 25

\bibitem[{{Koekemoer} {et~al.}(2011){Koekemoer}, {Faber}, {Ferguson}, {Grogin},
  {Kocevski}, {Koo}, {Lai}, {Lotz}, {Lucas}, {McGrath}, {Ogaz}, {Rajan},
  {Riess}, {Rodney}, {Strolger}, {Casertano}, {Castellano}, {Dahlen},
  {Dickinson}, {Dolch}, {Fontana}, {Giavalisco}, {Grazian}, {Guo}, {Hathi},
  {Huang}, {van der Wel}, {Yan}, {Acquaviva}, {Alexander}, {Almaini}, {Ashby},
  {Barden}, {Bell}, {Bournaud}, {Brown}, {Caputi}, {Cassata}, {Challis},
  {Chary}, {Cheung}, {Cirasuolo}, {Conselice}, {Roshan Cooray}, {Croton},
  {Daddi}, {Dav{\'e}}, {de Mello}, {de Ravel}, {Dekel}, {Donley}, {Dunlop},
  {Dutton}, {Elbaz}, {Fazio}, {Filippenko}, {Finkelstein}, {Frazer}, {Gardner},
  {Garnavich}, {Gawiser}, {Gruetzbauch}, {Hartley}, {H{\"a}ussler},
  {Herrington}, {Hopkins}, {Huang}, {Jha}, {Johnson}, {Kartaltepe},
  {Khostovan}, {Kirshner}, {Lani}, {Lee}, {Li}, {Madau}, {McCarthy},
  {McIntosh}, {McLure}, {McPartland}, {Mobasher}, {Moreira}, {Mortlock},
  {Moustakas}, {Mozena}, {Nandra}, {Newman}, {Nielsen}, {Niemi}, {Noeske},
  {Papovich}, {Pentericci}, {Pope}, {Primack}, {Ravindranath}, {Reddy},
  {Renzini}, {Rix}, {Robaina}, {Rosario}, {Rosati}, {Salimbeni}, {Scarlata},
  {Siana}, {Simard}, {Smidt}, {Snyder}, {Somerville}, {Spinrad}, {Straughn},
  {Telford}, {Teplitz}, {Trump}, {Vargas}, {Villforth}, {Wagner}, {Wandro},
  {Wechsler}, {Weiner}, {Wiklind}, {Wild}, {Wilson}, {Wuyts}, \&
  {Yun}}]{Koekemoer2011}
{Koekemoer}, A.~M., {Faber}, S.~M., {Ferguson}, H.~C., {et~al.} 2011, \apjs,
  197, 36

\bibitem[{{Kornei} {et~al.}(2013){Kornei}, {Shapley}, {Martin}, {Coil}, {Lotz},
  \& {Weiner}}]{Kornei2013}
{Kornei}, K.~A., {Shapley}, A.~E., {Martin}, C.~L., {et~al.} 2013, \apj, 774,
  50

\bibitem[{{Lange} {et~al.}(2016){Lange}, {van Dokkum}, {Momcheva}, {Nelson},
  {Leja}, {Brammer}, {Whitaker}, \& {Franx}}]{Lange2016}
{Lange}, J.~U., {van Dokkum}, P.~G., {Momcheva}, I.~G., {et~al.} 2016, \apjl,
  819, L4

\bibitem[{{Laor} {et~al.}(1997){Laor}, {Jannuzi}, {Green}, \&
  {Boroson}}]{Laor1997}
{Laor}, A., {Jannuzi}, B.~T., {Green}, R.~F., \& {Boroson}, T.~A. 1997, \apj,
  489, 656

\bibitem[{{Law} {et~al.}(2012){Law}, {Steidel}, {Shapley}, {Nagy}, {Reddy}, \&
  {Erb}}]{Law2012}
{Law}, D.~R., {Steidel}, C.~C., {Shapley}, A.~E., {et~al.} 2012, \apj, 759, 29

\bibitem[{{Leclercq} {et~al.}(2017){Leclercq}, {Bacon}, {Wisotzki}, {Mitchell},
  {Garel}, {Verhamme}, {Blaizot}, {Hashimoto}, {Herenz}, {Conseil},
  {Cantalupo}, {Inami}, {Contini}, {Richard}, {Maseda}, {Schaye}, {Marino},
  {Akhlaghi}, {Brinchmann}, \& {Carollo}}]{Leclercq2017}
{Leclercq}, F., {Bacon}, R., {Wisotzki}, L., {et~al.} 2017, \aap, 608, A8

\bibitem[{{Levesque} \& {Richardson}(2014)}]{Levesque2014}
{Levesque}, E.~M. \& {Richardson}, M.~L.~A. 2014, \apj, 780, 100

\bibitem[{{Lilly} {et~al.}(2013){Lilly}, {Carollo}, {Pipino}, {Renzini}, \&
  {Peng}}]{Lilly2013}
{Lilly}, S.~J., {Carollo}, C.~M., {Pipino}, A., {Renzini}, A., \& {Peng}, Y.
  2013, \apj, 772, 119

\bibitem[{{Luo} {et~al.}(2017){Luo}, {Brandt}, {Xue}, {Lehmer}, {Alexander},
  {Bauer}, {Vito}, {Yang}, {Basu-Zych}, {Comastri}, {Gilli}, {Gu},
  {Hornschemeier}, {Koekemoer}, {Liu}, {Mainieri}, {Paolillo}, {Ranalli},
  {Rosati}, {Schneider}, {Shemmer}, {Smail}, {Sun}, {Tozzi}, {Vignali}, \&
  {Wang}}]{Luo2017}
{Luo}, B., {Brandt}, W.~N., {Xue}, Y.~Q., {et~al.} 2017, \apjs, 228, 2

\bibitem[{{Madau} \& {Dickinson}(2014)}]{Madau2014}
{Madau}, P. \& {Dickinson}, M. 2014, \araa, 52, 415

\bibitem[{{Mainali} {et~al.}(2017){Mainali}, {Kollmeier}, {Stark}, {Simcoe},
  {Walth}, {Newman}, \& {Miller}}]{Mainali2017}
{Mainali}, R., {Kollmeier}, J.~A., {Stark}, D.~P., {et~al.} 2017, \apjl, 836,
  L14

\bibitem[{{Maiolino} {et~al.}(2008){Maiolino}, {Nagao}, {Grazian}, {Cocchia},
  {Marconi}, {Mannucci}, {Cimatti}, {Pipino}, {Ballero}, {Calura}, {Chiappini},
  {Fontana}, {Granato}, {Matteucci}, {Pastorini}, {Pentericci}, {Risaliti},
  {Salvati}, \& {Silva}}]{Maiolino2008}
{Maiolino}, R., {Nagao}, T., {Grazian}, A., {et~al.} 2008, \aap, 488, 463

\bibitem[{{Mannucci} {et~al.}(2009){Mannucci}, {Cresci}, {Maiolino}, {Marconi},
  {Pastorini}, {Pozzetti}, {Gnerucci}, {Risaliti}, {Schneider}, {Lehnert}, \&
  {Salvati}}]{Mannucci2009}
{Mannucci}, F., {Cresci}, G., {Maiolino}, R., {et~al.} 2009, \mnras, 398, 1915

\bibitem[{{Marino} {et~al.}(2018){Marino}, {Cantalupo}, {Lilly}, {Gallego},
  {Straka}, {Borisova}, {Pezzulli}, {Bacon}, {Brinchmann}, {Carollo},
  {Caruana}, {Conseil}, {Contini}, {Diener}, {Finley}, {Inami}, {Leclercq},
  {Muzahid}, {Richard}, {Schaye}, {Wendt}, \& {Wisotzki}}]{Marino2018}
{Marino}, R.~A., {Cantalupo}, S., {Lilly}, S.~J., {et~al.} 2018, \apj, 859, 53

\bibitem[{{Martin} {et~al.}(2012){Martin}, {Shapley}, {Coil}, {Kornei},
  {Bundy}, {Weiner}, {Noeske}, \& {Schiminovich}}]{Martin2012}
{Martin}, C.~L., {Shapley}, A.~E., {Coil}, A.~L., {et~al.} 2012, \apj, 760, 127

\bibitem[{{Martin} {et~al.}(2013){Martin}, {Shapley}, {Coil}, {Kornei},
  {Murray}, \& {Pancoast}}]{Martin2013}
{Martin}, C.~L., {Shapley}, A.~E., {Coil}, A.~L., {et~al.} 2013, \apj, 770, 41

\bibitem[{{Maseda} {et~al.}(2017){Maseda}, {Brinchmann}, {Franx}, {Bacon},
  {Bouwens}, {Schmidt}, {Boogaard}, {Contini}, {Feltre}, {Inami},
  {Kollatschny}, {Marino}, {Richard}, {Verhamme}, \& {Wisotzki}}]{Maseda2017}
{Maseda}, M.~V., {Brinchmann}, J., {Franx}, M., {et~al.} 2017, \aap, 608, A4

\bibitem[{{Matthee} {et~al.}(2017){Matthee}, {Sobral}, {Best}, {Khostovan},
  {Oteo}, {Bouwens}, \& {R{\"o}ttgering}}]{Matthee2017}
{Matthee}, J., {Sobral}, D., {Best}, P., {et~al.} 2017, \mnras, 465, 3637

\bibitem[{{Mentuch} {et~al.}(2010){Mentuch}, {Abraham}, \&
  {Zibetti}}]{Mentuch2010}
{Mentuch}, E., {Abraham}, R.~G., \& {Zibetti}, S. 2010, \apj, 725, 1971

\bibitem[{{Momcheva} {et~al.}(2016){Momcheva}, {Brammer}, {van Dokkum},
  {Skelton}, {Whitaker}, {Nelson}, {Fumagalli}, {Maseda}, {Leja}, {Franx},
  {Rix}, {Bezanson}, {Da Cunha}, {Dickey}, {F{\"o}rster Schreiber},
  {Illingworth}, {Kriek}, {Labb{\'e}}, {Ulf Lange}, {Lundgren}, {Magee},
  {Marchesini}, {Oesch}, {Pacifici}, {Patel}, {Price}, {Tal}, {Wake}, {van der
  Wel}, \& {Wuyts}}]{Momcheva2016}
{Momcheva}, I.~G., {Brammer}, G.~B., {van Dokkum}, P.~G., {et~al.} 2016, \apjs,
  225, 27

\bibitem[{{Nagao} {et~al.}(2006){Nagao}, {Maiolino}, \& {Marconi}}]{Nagao2006}
{Nagao}, T., {Maiolino}, R., \& {Marconi}, A. 2006, \aap, 459, 85

\bibitem[{{Nakajima} {et~al.}(2018{\natexlab{a}}){Nakajima}, {Fletcher},
  {Ellis}, {Robertson}, \& {Iwata}}]{Nakajima2018b}
{Nakajima}, K., {Fletcher}, T., {Ellis}, R.~S., {Robertson}, B.~E., \& {Iwata},
  I. 2018{\natexlab{a}}, \mnras, 477, 2098

\bibitem[{{Nakajima} {et~al.}(2018{\natexlab{b}}){Nakajima}, {Schaerer}, {Le
  F{\`e}vre}, {Amor{\'{\i}}n}, {Talia}, {Lemaux}, {Tasca}, {Vanzella},
  {Zamorani}, {Bardelli}, {Grazian}, {Guaita}, {Hathi}, {Pentericci}, \&
  {Zucca}}]{Nakajima2018a}
{Nakajima}, K., {Schaerer}, D., {Le F{\`e}vre}, O., {et~al.}
  2018{\natexlab{b}}, \aap, 612, A94

\bibitem[{{Newman} {et~al.}(2013){Newman}, {Cooper}, {Davis}, {Faber}, {Coil},
  {Guhathakurta}, {Koo}, {Phillips}, {Conroy}, {Dutton}, {Finkbeiner}, {Gerke},
  {Rosario}, {Weiner}, {Willmer}, {Yan}, {Harker}, {Kassin}, {Konidaris},
  {Lai}, {Madgwick}, {Noeske}, {Wirth}, {Connolly}, {Kaiser}, {Kirby},
  {Lemaux}, {Lin}, {Lotz}, {Luppino}, {Marinoni}, {Matthews}, {Metevier}, \&
  {Schiavon}}]{Newman2013}
{Newman}, J.~A., {Cooper}, M.~C., {Davis}, M., {et~al.} 2013, \apjs, 208, 5

\bibitem[{{Nilsson} {et~al.}(2009){Nilsson}, {M{\"o}ller-Nilsson}, {M{\o}ller},
  {Fynbo}, \& {Shapley}}]{Nilsson2009}
{Nilsson}, K.~K., {M{\"o}ller-Nilsson}, O., {M{\o}ller}, P., {Fynbo}, J.~P.~U.,
  \& {Shapley}, A.~E. 2009, \mnras, 400, 232

\bibitem[{{Panuzzo} {et~al.}(2003){Panuzzo}, {Bressan}, {Granato}, {Silva}, \&
  {Danese}}]{Panuzzo2003}
{Panuzzo}, P., {Bressan}, A., {Granato}, G.~L., {Silva}, L., \& {Danese}, L.
  2003, \aap, 409, 99

\bibitem[{{Paulino-Afonso} {et~al.}(2018){Paulino-Afonso}, {Sobral}, {Ribeiro},
  {Matthee}, {Santos}, {Calhau}, {Forshaw}, {Johnson}, {Merrick}, {P{\'e}rez},
  \& {Sheldon}}]{Paulino-Afonso2018}
{Paulino-Afonso}, A., {Sobral}, D., {Ribeiro}, B., {et~al.} 2018, \mnras, 476,
  5479

\bibitem[{{Peng} {et~al.}(2010){Peng}, {Ho}, {Impey}, \& {Rix}}]{Peng2010}
{Peng}, C.~Y., {Ho}, L.~C., {Impey}, C.~D., \& {Rix}, H.-W. 2010, \aj, 139,
  2097

\bibitem[{{Planck Collaboration} {et~al.}(2016){Planck Collaboration}, {Ade},
  {Aghanim}, {Arnaud}, {Ashdown}, {Aumont}, {Baccigalupi}, {Banday},
  {Barreiro}, {Bartlett}, \& et~al.}]{Planck2016}
{Planck Collaboration}, {Ade}, P.~A.~R., {Aghanim}, N., {et~al.} 2016, \aap,
  594, A13

\bibitem[{{Prochaska} {et~al.}(2011){Prochaska}, {Kasen}, \&
  {Rubin}}]{Prochaska2011}
{Prochaska}, J.~X., {Kasen}, D., \& {Rubin}, K. 2011, \apj, 734, 24

\bibitem[{{Rafelski} {et~al.}(2015){Rafelski}, {Teplitz}, {Gardner}, {Coe},
  {Bond}, {Koekemoer}, {Grogin}, {Kurczynski}, {McGrath}, {Bourque}, {Atek},
  {Brown}, {Colbert}, {Codoreanu}, {Ferguson}, {Finkelstein}, {Gawiser},
  {Giavalisco}, {Gronwall}, {Hanish}, {Lee}, {Mehta}, {de Mello},
  {Ravindranath}, {Ryan}, {Scarlata}, {Siana}, {Soto}, \&
  {Voyer}}]{Rafelski2015}
{Rafelski}, M., {Teplitz}, H.~I., {Gardner}, J.~P., {et~al.} 2015, \aj, 150, 31

\bibitem[{{Rich} {et~al.}(2011){Rich}, {Kewley}, \& {Dopita}}]{Rich2011}
{Rich}, J.~A., {Kewley}, L.~J., \& {Dopita}, M.~A. 2011, \apj, 734, 87

\bibitem[{{Rich} {et~al.}(2014){Rich}, {Kewley}, \& {Dopita}}]{Rich2014}
{Rich}, J.~A., {Kewley}, L.~J., \& {Dopita}, M.~A. 2014, \apjl, 781, L12

\bibitem[{{Rigby} {et~al.}(2014){Rigby}, {Bayliss}, {Gladders}, {Sharon},
  {Wuyts}, \& {Dahle}}]{Rigby2014}
{Rigby}, J.~R., {Bayliss}, M.~B., {Gladders}, M.~D., {et~al.} 2014, \apj, 790,
  44

\bibitem[{{Rodighiero} {et~al.}(2007){Rodighiero}, {Cimatti}, {Franceschini},
  {Brusa}, {Fritz}, \& {Bolzonella}}]{Rodighiero2007}
{Rodighiero}, G., {Cimatti}, A., {Franceschini}, A., {et~al.} 2007, \aap, 470,
  21

\bibitem[{{Rubin} {et~al.}(2011){Rubin}, {Prochaska}, {M{\'e}nard}, {Murray},
  {Kasen}, {Koo}, \& {Phillips}}]{Rubin2011}
{Rubin}, K.~H.~R., {Prochaska}, J.~X., {M{\'e}nard}, B., {et~al.} 2011, \apj,
  728, 55

\bibitem[{{Rubin} {et~al.}(2010){Rubin}, {Weiner}, {Koo}, {Martin},
  {Prochaska}, {Coil}, \& {Newman}}]{Rubin2010}
{Rubin}, K.~H.~R., {Weiner}, B.~J., {Koo}, D.~C., {et~al.} 2010, \apj, 719,
  1503

\bibitem[{{S{\'a}nchez-Bl{\'a}zquez} {et~al.}(2006){S{\'a}nchez-Bl{\'a}zquez},
  {Peletier}, {Jim{\'e}nez-Vicente}, {Cardiel}, {Cenarro},
  {Falc{\'o}n-Barroso}, {Gorgas}, {Selam}, \&
  {Vazdekis}}]{Sanchez-Blazquez2006}
{S{\'a}nchez-Bl{\'a}zquez}, P., {Peletier}, R.~F., {Jim{\'e}nez-Vicente}, J.,
  {et~al.} 2006, \mnras, 371, 703

\bibitem[{{Scarlata} \& {Panagia}(2015)}]{Scarlata2015}
{Scarlata}, C. \& {Panagia}, N. 2015, \apj, 801, 43

\bibitem[{{Schaerer} {et~al.}(2016){Schaerer}, {Izotov}, {Verhamme},
  {Orlitov{\'a}}, {Thuan}, {Worseck}, \& {Guseva}}]{Schaerer2016}
{Schaerer}, D., {Izotov}, Y.~I., {Verhamme}, A., {et~al.} 2016, \aap, 591, L8

\bibitem[{{Senchyna} {et~al.}(2017){Senchyna}, {Stark}, {Vidal-Garc{\'{\i}}a},
  {Chevallard}, {Charlot}, {Mainali}, {Jones}, {Wofford}, {Feltre}, \&
  {Gutkin}}]{Senchyna2017}
{Senchyna}, P., {Stark}, D.~P., {Vidal-Garc{\'{\i}}a}, A., {et~al.} 2017,
  \mnras, 472, 2608

\bibitem[{{Shen} {et~al.}(2003){Shen}, {Mo}, {White}, {Blanton}, {Kauffmann},
  {Voges}, {Brinkmann}, \& {Csabai}}]{Shen2003}
{Shen}, S., {Mo}, H.~J., {White}, S.~D.~M., {et~al.} 2003, \mnras, 343, 978

\bibitem[{{Shibuya} {et~al.}(2014){Shibuya}, {Ouchi}, {Nakajima}, {Yuma},
  {Hashimoto}, {Shimasaku}, {Mori}, \& {Umemura}}]{Shibuya2014}
{Shibuya}, T., {Ouchi}, M., {Nakajima}, K., {et~al.} 2014, \apj, 785, 64

\bibitem[{{Shirazi} \& {Brinchmann}(2012)}]{Shirazi2012}
{Shirazi}, M. \& {Brinchmann}, J. 2012, \mnras, 421, 1043

\bibitem[{{Shivaei} {et~al.}(2018){Shivaei}, {Reddy}, {Siana}, {Shapley},
  {Kriek}, {Mobasher}, {Freeman}, {Sanders}, {Coil}, {Price}, {Fetherolf},
  {Azadi}, {Leung}, \& {Zick}}]{Shivaei2018}
{Shivaei}, I., {Reddy}, N.~A., {Siana}, B., {et~al.} 2018, \apj, 855, 42

\bibitem[{{Stark}(2016)}]{Stark2016}
{Stark}, D.~P. 2016, \araa, 54, 761

\bibitem[{{Stark} {et~al.}(2017){Stark}, {Ellis}, {Charlot}, {Chevallard},
  {Tang}, {Belli}, {Zitrin}, {Mainali}, {Gutkin}, {Vidal-Garc{\'{\i}}a},
  {Bouwens}, \& {Oesch}}]{Stark2017}
{Stark}, D.~P., {Ellis}, R.~S., {Charlot}, S., {et~al.} 2017, \mnras, 464, 469

\bibitem[{{Stark} {et~al.}(2010){Stark}, {Ellis}, {Chiu}, {Ouchi}, \&
  {Bunker}}]{Stark2010}
{Stark}, D.~P., {Ellis}, R.~S., {Chiu}, K., {Ouchi}, M., \& {Bunker}, A. 2010,
  \mnras, 408, 1628

\bibitem[{{Stark} {et~al.}(2015{\natexlab{a}}){Stark}, {Richard}, {Charlot},
  {Cl{\'e}ment}, {Ellis}, {Siana}, {Robertson}, {Schenker}, {Gutkin}, \&
  {Wofford}}]{Stark2015a}
{Stark}, D.~P., {Richard}, J., {Charlot}, S., {et~al.} 2015{\natexlab{a}},
  \mnras, 450, 1846

\bibitem[{{Stark} {et~al.}(2014){Stark}, {Richard}, {Siana}, {Charlot},
  {Freeman}, {Gutkin}, {Wofford}, {Robertson}, {Amanullah}, {Watson}, \&
  {Milvang-Jensen}}]{Stark2014}
{Stark}, D.~P., {Richard}, J., {Siana}, B., {et~al.} 2014, \mnras, 445, 3200

\bibitem[{{Stark} {et~al.}(2015{\natexlab{b}}){Stark}, {Walth}, {Charlot},
  {Cl{\'e}ment}, {Feltre}, {Gutkin}, {Richard}, {Mainali}, {Robertson},
  {Siana}, {Tang}, \& {Schenker}}]{Stark2015b}
{Stark}, D.~P., {Walth}, G., {Charlot}, S., {et~al.} 2015{\natexlab{b}},
  \mnras, 454, 1393

\bibitem[{{Stasi{\'n}ska} \& {Leitherer}(1996)}]{Stasinska1996}
{Stasi{\'n}ska}, G. \& {Leitherer}, C. 1996, \apjs, 107, 661

\bibitem[{{Steidel} {et~al.}(2010){Steidel}, {Erb}, {Shapley}, {Pettini},
  {Reddy}, {Bogosavljevi{\'c}}, {Rudie}, \& {Rakic}}]{Steidel2010}
{Steidel}, C.~C., {Erb}, D.~K., {Shapley}, A.~E., {et~al.} 2010, \apj, 717, 289

\bibitem[{{Steidel} {et~al.}(2004){Steidel}, {Shapley}, {Pettini},
  {Adelberger}, {Erb}, {Reddy}, \& {Hunt}}]{Steidel2004}
{Steidel}, C.~C., {Shapley}, A.~E., {Pettini}, M., {et~al.} 2004, \apj, 604,
  534

\bibitem[{{Tremonti} {et~al.}(2004){Tremonti}, {Heckman}, {Kauffmann},
  {Brinchmann}, {Charlot}, {White}, {Seibert}, {Peng}, {Schlegel}, {Uomoto},
  {Fukugita}, \& {Brinkmann}}]{Tremonti2004}
{Tremonti}, C.~A., {Heckman}, T.~M., {Kauffmann}, G., {et~al.} 2004, \apj, 613,
  898

\bibitem[{{Tremonti} {et~al.}(2007){Tremonti}, {Moustakas}, \&
  {Diamond-Stanic}}]{Tremonti2007}
{Tremonti}, C.~A., {Moustakas}, J., \& {Diamond-Stanic}, A.~M. 2007, \apjl,
  663, L77

\bibitem[{{van der Wel} {et~al.}(2012){van der Wel}, {Bell}, {H{\"a}ussler},
  {McGrath}, {Chang}, {Guo}, {McIntosh}, {Rix}, {Barden}, {Cheung}, {Faber},
  {Ferguson}, {Galametz}, {Grogin}, {Hartley}, {Kartaltepe}, {Kocevski},
  {Koekemoer}, {Lotz}, {Mozena}, {Peth}, \& {Peng}}]{VanderWel2012}
{van der Wel}, A., {Bell}, E.~F., {H{\"a}ussler}, B., {et~al.} 2012, \apjs,
  203, 24

\bibitem[{{van der Wel} {et~al.}(2014){van der Wel}, {Franx}, {van Dokkum},
  {Skelton}, {Momcheva}, {Whitaker}, {Brammer}, {Bell}, {Rix}, {Wuyts},
  {Ferguson}, {Holden}, {Barro}, {Koekemoer}, {Chang}, {McGrath},
  {H{\"a}ussler}, {Dekel}, {Behroozi}, {Fumagalli}, {Leja}, {Lundgren},
  {Maseda}, {Nelson}, {Wake}, {Patel}, {Labb{\'e}}, {Faber}, {Grogin}, \&
  {Kocevski}}]{VanderWel2014}
{van der Wel}, A., {Franx}, M., {van Dokkum}, P.~G., {et~al.} 2014, \apj, 788,
  28

\bibitem[{{Vanzella} {et~al.}(2016){Vanzella}, {De Barros}, {Cupani}, {Karman},
  {Gronke}, {Balestra}, {Coe}, {Mignoli}, {Brusa}, {Calura}, {Caminha},
  {Caputi}, {Castellano}, {Christensen}, {Comastri}, {Cristiani}, {Dijkstra},
  {Fontana}, {Giallongo}, {Giavalisco}, {Gilli}, {Grazian}, {Grillo},
  {Koekemoer}, {Meneghetti}, {Nonino}, {Pentericci}, {Rosati}, {Schaerer},
  {Verhamme}, {Vignali}, \& {Zamorani}}]{Vanzella2016}
{Vanzella}, E., {De Barros}, S., {Cupani}, G., {et~al.} 2016, \apjl, 821, L27

\bibitem[{{Veilleux} {et~al.}(2005){Veilleux}, {Cecil}, \&
  {Bland-Hawthorn}}]{Veilleux2005}
{Veilleux}, S., {Cecil}, G., \& {Bland-Hawthorn}, J. 2005, \araa, 43, 769

\bibitem[{{Ventou} {et~al.}(2017){Ventou}, {Contini}, {Bouch{\'e}}, {Epinat},
  {Brinchmann}, {Bacon}, {Inami}, {Lam}, {Drake}, {Garel}, {Michel-Dansac},
  {Pello}, {Steinmetz}, {Weilbacher}, {Wisotzki}, \& {Carollo}}]{Ventou2017}
{Ventou}, E., {Contini}, T., {Bouch{\'e}}, N., {et~al.} 2017, \aap, 608, A9

\bibitem[{{Verhamme} {et~al.}(2008){Verhamme}, {Schaerer}, {Atek}, \&
  {Tapken}}]{Verhamme2008}
{Verhamme}, A., {Schaerer}, D., {Atek}, H., \& {Tapken}, C. 2008, \aap, 491, 89

\bibitem[{{Vidal-Garc{\'{\i}}a} {et~al.}(2017){Vidal-Garc{\'{\i}}a}, {Charlot},
  {Bruzual}, \& {Hubeny}}]{Vidal2017}
{Vidal-Garc{\'{\i}}a}, A., {Charlot}, S., {Bruzual}, G., \& {Hubeny}, I. 2017,
  \mnras, 470, 3532

\bibitem[{{Weiner} {et~al.}(2009){Weiner}, {Coil}, {Prochaska}, {Newman},
  {Cooper}, {Bundy}, {Conselice}, {Dutton}, {Faber}, {Koo}, {Lotz}, {Rieke}, \&
  {Rubin}}]{Weiner2009}
{Weiner}, B.~J., {Coil}, A.~L., {Prochaska}, J.~X., {et~al.} 2009, \apj, 692,
  187

\bibitem[{{Whitaker} {et~al.}(2014){Whitaker}, {Franx}, {Leja}, {van Dokkum},
  {Henry}, {Skelton}, {Fumagalli}, {Momcheva}, {Brammer}, {Labb{\'e}},
  {Nelson}, \& {Rigby}}]{Whitaker2014}
{Whitaker}, K.~E., {Franx}, M., {Leja}, J., {et~al.} 2014, \apj, 795, 104

\bibitem[{{Wisotzki} {et~al.}(2016){Wisotzki}, {Bacon}, {Blaizot},
  {Brinchmann}, {Herenz}, {Schaye}, {Bouch{\'e}}, {Cantalupo}, {Contini},
  {Carollo}, {Caruana}, {Courbot}, {Emsellem}, {Kamann}, {Kerutt}, {Leclercq},
  {Lilly}, {Patr{\'{\i}}cio}, {Sandin}, {Steinmetz}, {Straka}, {Urrutia},
  {Verhamme}, {Weilbacher}, \& {Wendt}}]{Wisotzki2016}
{Wisotzki}, L., {Bacon}, R., {Blaizot}, J., {et~al.} 2016, \aap, 587, A98

\bibitem[{{Wofford} {et~al.}(2016){Wofford}, {Charlot}, {Bruzual}, {Eldridge},
  {Calzetti}, {Adamo}, {Cignoni}, {de Mink}, {Gouliermis}, {Grasha}, {Grebel},
  {Lee}, {{\"O}stlin}, {Smith}, {Ubeda}, \& {Zackrisson}}]{Wofford2016}
{Wofford}, A., {Charlot}, S., {Bruzual}, G., {et~al.} 2016, \mnras, 457, 4296

\bibitem[{{York} {et~al.}(2000){York}, {Adelman}, {Anderson}, {Anderson},
  {Annis}, {Bahcall}, {Bakken}, {Barkhouser}, {Bastian}, {Berman}, {Boroski},
  {Bracker}, {Briegel}, {Briggs}, {Brinkmann}, {Brunner}, {Burles}, {Carey},
  {Carr}, {Castander}, {Chen}, {Colestock}, {Connolly}, {Crocker}, {Csabai},
  {Czarapata}, {Davis}, {Doi}, {Dombeck}, {Eisenstein}, {Ellman}, {Elms},
  {Evans}, {Fan}, {Federwitz}, {Fiscelli}, {Friedman}, {Frieman}, {Fukugita},
  {Gillespie}, {Gunn}, {Gurbani}, {de Haas}, {Haldeman}, {Harris}, {Hayes},
  {Heckman}, {Hennessy}, {Hindsley}, {Holm}, {Holmgren}, {Huang}, {Hull},
  {Husby}, {Ichikawa}, {Ichikawa}, {Ivezi{\'c}}, {Kent}, {Kim}, {Kinney},
  {Klaene}, {Kleinman}, {Kleinman}, {Knapp}, {Korienek}, {Kron}, {Kunszt},
  {Lamb}, {Lee}, {Leger}, {Limmongkol}, {Lindenmeyer}, {Long}, {Loomis},
  {Loveday}, {Lucinio}, {Lupton}, {MacKinnon}, {Mannery}, {Mantsch}, {Margon},
  {McGehee}, {McKay}, {Meiksin}, {Merelli}, {Monet}, {Munn}, {Narayanan},
  {Nash}, {Neilsen}, {Neswold}, {Newberg}, {Nichol}, {Nicinski}, {Nonino},
  {Okada}, {Okamura}, {Ostriker}, {Owen}, {Pauls}, {Peoples}, {Peterson},
  {Petravick}, {Pier}, {Pope}, {Pordes}, {Prosapio}, {Rechenmacher}, {Quinn},
  {Richards}, {Richmond}, {Rivetta}, {Rockosi}, {Ruthmansdorfer}, {Sandford},
  {Schlegel}, {Schneider}, {Sekiguchi}, {Sergey}, {Shimasaku}, {Siegmund},
  {Smee}, {Smith}, {Snedden}, {Stone}, {Stoughton}, {Strauss}, {Stubbs},
  {SubbaRao}, {Szalay}, {Szapudi}, {Szokoly}, {Thakar}, {Tremonti}, {Tucker},
  {Uomoto}, {Vanden Berk}, {Vogeley}, {Waddell}, {Wang}, {Watanabe},
  {Weinberg}, {Yanny}, {Yasuda}, \& {SDSS Collaboration}}]{York2000}
{York}, D.~G., {Adelman}, J., {Anderson}, Jr., J.~E., {et~al.} 2000, \aj, 120,
  1579

\bibitem[{{Yuan} {et~al.}(2012){Yuan}, {Kewley}, {Swinbank}, \&
  {Richard}}]{Yuan2012}
{Yuan}, T.-T., {Kewley}, L.~J., {Swinbank}, A.~M., \& {Richard}, J. 2012, \apj,
  759, 66

\bibitem[{{Zhu} {et~al.}(2015){Zhu}, {Comparat}, {Kneib}, {Delubac},
  {Raichoor}, {Dawson}, {Newman}, {Y{\`e}che}, {Zhou}, \&
  {Schneider}}]{Zhu2015}
{Zhu}, G.~B., {Comparat}, J., {Kneib}, J.-P., {et~al.} 2015, \apj, 815, 48

\end{thebibliography}

\end{document}